%% file: apssamp.tex
\documentclass[
aps,
prl,
showpacs,
letterpaper,
twocolumn,
superscriptaddress,
notitlepage,
nofootinbib,
floatfix
]{revtex4-2}

\usepackage{graphicx}
\usepackage{dcolumn}
\usepackage{bm}
\usepackage{aas_macros}

\usepackage{xspace}
\usepackage{longtable}
\usepackage{orcidlink}

\newcommand{\mosfit}{\texttt{MOSFiT}\xspace}
\newcommand{\Msol}{\mbox{$M_{\odot}$}}

\DeclareRobustCommand{\ion}[2]{%
\relax\ifmmode
\ifx\testbx\f@series
{\mathbf{#1\,\mathsc{#2}}}\else
{\mathrm{#1\,\mathsc{#2}}}\fi
\else\textup{#1\,{\mdseries\textsc{#2}}}%
\fi}

\begin{document}


\title{SN 2023uqf: An Interacting Supernova Coincident with a High-Energy Neutrino}

\input{authors}








\date{\today}

\begin{abstract}
Astrophysical high-energy (TeV-PeV) neutrinos were first discovered in 2013, but their origin remains largely unknown. Here we present SN 2023uqf, a supernova found in coincidence with high-energy neutrino IC231004A, as part of a systematic optical follow-up program with the Zwicky Transient Facility. SN 2023uqf had a luminous and rapidly-evolving lightcurve, and spectroscopic observations indicated that the source was a Type Ibn supernova. Spectroscopic signatures confirm ongoing interaction between the supernova ejecta and a dense circumstellar medium, as expected for high-energy neutrino production in a core-collapse supernova. 
Given the rare nature of Type Ibn supernovae, SN 2023uqf is unlikely to have been discovered by chance over the course of our program (p$=0.3\%$). 
Our discovery of SN 2023uqf provides the first observational evidence to support long-held theories that interacting supernovae can serve as cosmic hadron accelerators. 

\end{abstract}

\maketitle


Supernova 1987A launched the field of multi-messenger astrophysics, as a supernova in the nearby Large Magellanic Cloud for which coincident MeV neutrino emission was also detected \cite{sn1987a_neutrino}. A new `era of multi-messenger astronomy' has emerged in the last decade, with the discovery of high-energy (TeV-PeV) astrophysical neutrinos by IceCube in 2013 \cite{ic_astro}, and gravitational waves by LIGO in 2015 \cite{ligo_bbh_16}. IceCube has reported that some high-energy neutrinos are emitted by our own galaxy \cite{ic_gp_23}, as well as by nearby active galaxy NGC 1068 \cite{ic_ps_10_yr,ic_ngc_22}, flaring blazar TXS 0506+056 \cite{txs_mm,txs_ps}. There is also evidence for neutrino emission from the wider blazar population \citep{padovani_16,plavin_20,hovatta_20,buson_22}, and tidal disruption events \cite{bran,tywin,lancel,jiang_23}. However, much of the flux remains unaccounted for. Gravitational waves have now been detected from the coalescence of compact objects (pairs of neutron stars and black holes \cite{ligo_bbh_16,gw170817_bns,ligo_nsbh_21}), and the  electromagnetic counterpart to one binary neutron star merger (GW170817) has been detected \cite{gw170818_mm}, as have several potential counterparts to binary black hole mergers \citep{graham_20_bbh, graham_23}. 

Core-collapse supernovae (CCSNe) have been conspicuous by their absence in this new era. CCSNe, the explosive deaths of massive stars, are luminous astrophysical transients typically discovered via optical emission. CCSNe are also prolific non-thermal emitters, detected in radio \citep{weiler_2002, bietenholz_21}, and launching outflows \citep{paragi_10}. It has long been known that the rare subclass of Type Ic broad-lined CCSNe can launch short-lived relativistic particle jets known as `Gamma-Ray Bursts' (GRBs) \citep{galama,Kulkarni}, from which photons with energies exceeding 1 TeV have been detected \cite{magic_grb191114c,hess_grb180720b,lhaaso_grb221009a}. More recently, the first high-confidence detection of persistent gamma-rays from a supernova was reported \citep{chen_24}. SN 2022jli, an unusual stripped-envelope CCSN suspected to have a compact companion, exhibited a relatively brief ($\sim$60 day) but luminous (L$_{\gamma} > 3 \times 10^{41}$ erg s$^{-1}$) gamma-ray flare peaking at energies of 1--3 GeV \citep{chen_24}. These observations confirm that at least some CCSNe are efficient accelerators of electrons. 

Perhaps unsurprisingly, it has been predicted that a substantial fraction of extragalactic high-energy neutrinos and cosmic rays should also arise from the abundant core-collapse supernova population. These models generally consider either relativistic jets \cite{meszaros_01, soebur_05, horiuchi_08,  murase_13, senno_16, ning_18} or the interaction of CCSN ejecta with circumstellar material (CSM) \cite{murase_11,zirakashvili_16,petropolou_17,murase_18,murase_19,wang_19,sarmah_22,murase_24,waxman_24}. Some focus specifically on `choked jets', in which a relativistic jet fails to escape the stellar envelope or surrounding CSM, leading to neutrino emission but no other high-energy signatures \citep{meszaros_01, murase_13, nakar_15,senno_16}.

However, archival searches between known core-collapse supernovae and neutrinos have not revealed any significant correlations \citep{ic_sn_23,chang_24}. Similarly, no correlation between long GRBs and neutrinos has been found despite a decade of searches \citep{ic_15_grb,ic_22_grb}, and the lack of neutrinos associated to the recent bright GRB 221009A further confirmed that only a tiny fraction ($\lesssim$ 1\%) of neutrinos can originate from GRBs \citep{ic_grb221009a, murase_22,veres_24}. 

A complementary approach to identify neutrino-emitting supernovae is via dedicated neutrino follow-up programs with optical telescopes \citep{kowalski_07}. Neutrino production models have associated temporal and spectroscopic signatures which would be expected to accompany neutrino production, and this enables the vast majority of optical transients to be rejected. In particular, neutrino production via CSM should be accompanied by clearly identifiable signatures of interaction in the optical spectra (see \citet{smith_17} for a recent review). IceCube reports all likely high-energy neutrinos via public realtime alerts \citep{ic_realtime,icecat1}, with typical 90\% localisation areas of $\sim5$ -- 10 sq. degrees over the course of our program\footnote{As of September 2024, IceCube has transitioned to a new reconstruction approach with average localisation areas that are closer to 1 sq. deg. \citep{ic_gcn_reco}.}. These neutrino alerts are systematically followed-up by instruments including ASAS-SN, DECam, Pan-STARRS1, and ZTF \citep{asassn_neutrino,decam_neutrino,panstarrs_neutrino,stein_23}. Candidate supernova-neutrino associations have in fact been tentatively identified in the past by these programs, but all candidates could ultimately be ruled out by subsequent multi-wavelength observations \cite{ptf_neutrino,panstarrs_neutrino,stein_23}.

In particular, ZTF \citep{bellm_19,graham_19,masci_19,dekaney_20,ampel_19,skyportal_19,skyportal_23} has operated a long-running neutrino follow-up program since March 2018 \citep{stein_23}, comprising 43 individual campaigns as of September 2024 (see SM for further details). Optical transients and variable sources found in these campaigns are systematically identified and classified, yielding a comprehensive census of neutrino-coincident sources. From these 43 campaigns, which tiled a cumulative 257 sq. deg. of sky, two TDEs were identified as probable neutrino sources \cite{bran, lancel}. The ZTF follow-up program, by virtue of its depth and duration, is the most sensitive optical follow-up program ever conducted for transients such as supernovae \citep{stein_23}. 

\subsection{Observations of SN 2023uqf}
IC231004A, the 38th neutrino alert followed up by ZTF, was detected by IceCube on 2023 October 4 at 14:39:41.18 UT with an unusually high estimated energy (442 TeV) and probability of being astrophysical (84\%) \cite{ic_gcn_1, ic_gcn_2}. The reported 90\% localisation region was 4.3 sq. deg. \cite{ic_gcn_1, ic_gcn_2}, of which 3.1 sq. deg. were tiled by ZTF observations beginning 21.3 hours after neutrino detection. Accounting for both the astrophysical probability and our ZTF coverage, the probability that an astrophysical counterpart lay within the ZTF footprint was 61\%, which is higher than a typical neutrino in our program \citep{stein_23}. After analysing data to identify possible counterparts (see \citet{stein_23} for more details), only one candidate transient was found.

ZTF23abidzvf, first discovered by ZTF in our ToO observations, was reported to the Transient Name Server as an astronomical transient and assigned the designation 2023uqf \cite{tns_disc}. The candidate was published as a possible neutrino counterpart via NASA's General Coordinates Network (GCN) \citep{gcn_disc}. As seen in Figure \ref{fig:mosfit_lc}, the source evolved very rapidly in the days after discovery, brightening by $\sim$1 magnitude in 2 days and then fading rapidly. This extreme behaviour clearly marked out SN 2023uqf as an interesting candidate counterpart, and prompted additional photometric observation (see SM for details).

\begin{figure}
    \centering
    \includegraphics[width=0.45\textwidth]{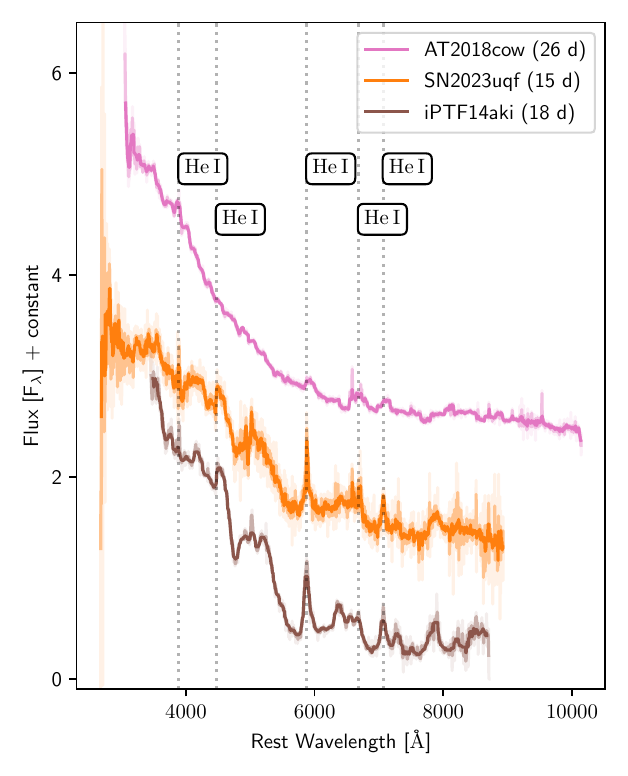}
    \caption{Spectrum of SN 2023uqf, alongside Type Ibn supernova iPTF14aki \cite{Hosseinzadeh2017} and LFBOT AT 2018cow \citep{perley_at2018cow}. The time of each spectrum is given relative to first optical detection. The vertical dashed lines indicate the He I lines, which are clearly detected for SN 2023uqf and iPTF14aki. Given the resemblance between the two, we classify SN 2023uqf as a Type Ibn supernova \citep{gcn_class,tns_class} }
    \label{fig:Spectrum}
\end{figure}

Due to moon proximity, spectroscopic observations were first obtained 13 days after neutrino detection, on 2023-10-17 with ALFOSC at NOT and LRIS at Keck-I (see SM for further details). The spectra show a hot and mostly featureless blue continuum with numerous narrow host-galaxy emission features (H$\alpha$, H$\beta$, [\ion{O}{III}], [\ion{O}{II}]) at a consistent redshift of $z=0.15$. As seen in Figure \ref{fig:Spectrum}, there are intermediate-width emission features consistent with redshifted \ion{He}{I} also visible (including those at rest-frame wavelengths of~4471$\rm~\AA$ and~5876$\rm~\AA$). No other individually-identifiable features from the transient were observed. We highlight in particular the lack of any similarly-wide H lines in the spectrum (only narrow lines from the host galaxy are present). Based on these observations, and the general similarity of the spectra to Type Ibn events from the literature such as iPTF14aki \citep{Hosseinzadeh2017} at a similar rest-frame epoch, we classified SN 2023uqf as a Type Ibn supernova \citep{gcn_class,tns_class} (see also Fig.~\ref{fig:Spectrum}). The subsequent spectroscopic evolution of SN 2023uqf was consistent with this classification, including the characteristic fading of P-Cygni line profiles with late-time blueshifting (see also SM Figure 7).

The narrow He lines present in the spectra of SN 2023uqf confirm the presence of CSM interaction in this source. SN 2023uqf thus unambiguously exhibits all the expected spectroscopic and temporal behaviour expected for the popular CSM neutrino production models \cite{murase_11,zirakashvili_16,petropolou_17,murase_18,murase_19,wang_19,sarmah_22,murase_24,waxman_24}. Type Ibn SNe are rare (representing $\sim$1\% of the overall CCSNe rate \citep{maeda_22}). Our spectra show that SN 2023uqf, like all Type Ibn SNe, experienced substantial mass loss prior to explosion with the hydrogen shell being entirely stripped. After explosion, the supernova shock collided with a dense He-rich/H-poor CSM. These observations confirmed that SN 2023uqf had characteristics expected of a multi-messenger supernova. 

\begin{figure}
    \centering
    \includegraphics[width=0.5\textwidth]{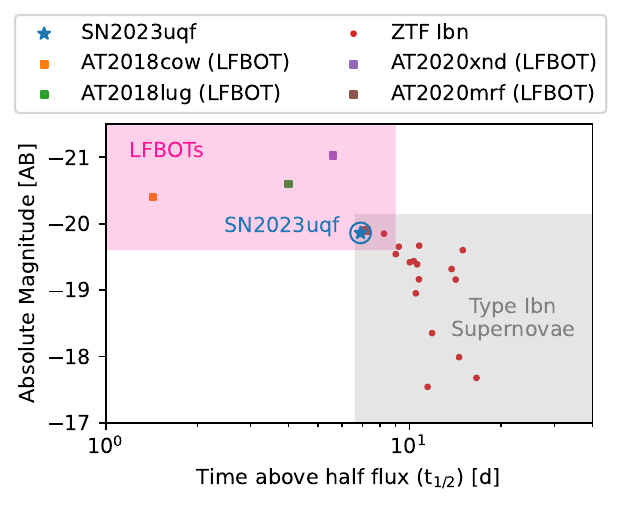}
    \caption{Rest-frame time above half peak flux (t$_{1/2}$) versus peak magnitude for a flux-limited sample of Type Ibn SNe from ZTF \citep{Gangopadhyay2025} (brown circles) and a selection of LFBOTs from the literature \cite{at2018lug,at2020xnd,perley_at2018cow,yao_22_mrf} (squares). SN 2023uqf (blue star) lies in the overlap region between these populations, close to AT 2020mrf. See SM for further details.}
    \label{fig:duration_vs_peak}
\end{figure}

\begin{figure}
    \centering
    \includegraphics[width=0.48\textwidth]{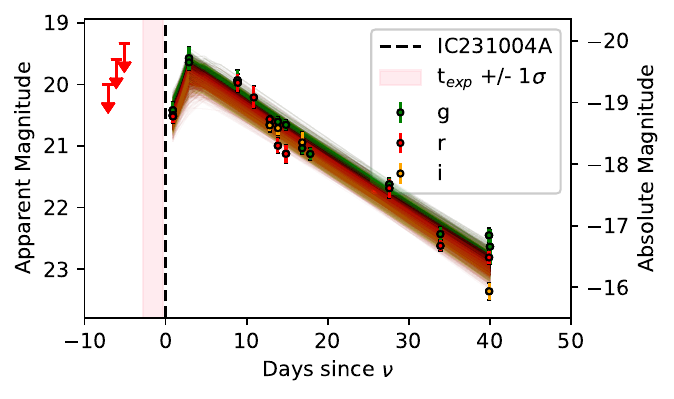}
    \caption{CSM-interaction fit of the SN 2023uqf lightcurve (see SM for further details). The time of detection of IC231004A is indicated with the dashed line, alongside the estimated explosion window. Upper limits are shown with arrows.}
    \label{fig:mosfit_lc}
\end{figure}

The photometric and spectroscopic evolution of SN 2023uqf were unusual for a SN Ibn. As seen in Figure \ref{fig:mosfit_lc}, though the exact peak of the supernova light curve was not resolved, the redshift measurement confirms the source reached an absolute $g-$band magnitude of at least M$_{g} = -19.7$ ($\nu L_{\nu} \approx 2 \times 10^{43}$ erg s$^{-1}$). Even this value would place SN 2023uqf among the more luminous Type Ibn supernovae, as seen in Figure \ref{fig:duration_vs_peak}. In addition, the source evolved rapidly, with a t$_{1/2}$ duration of just 8 days. As seen in Figure \ref{fig:duration_vs_peak}, this places SN 2023uqf as the most rapid of all ZTF Type Ibn supernovae, and in the region of known overlap between extreme Type Ibn SNe and `luminous fast blue optical transients' (LFBOTs) \citep{shivvers_16,fox_2019,ho_fbot_sample_23,Pellegrino2022,maeda_22}. In particular, the $g$-band evolution of SN 2023uqf bears a striking resemblance to the LFBOT AT 2020mrf \citep{yao_22_mrf} (see also SM Figure 3).
LFBOTs are a population of astrophysical transients characterised by optical emission which is typically luminous (M $\lesssim -20$), rapidly evolving (t$_{1/2} \lesssim 10$d), and blue (hot thermal emission with T $\approx$ 20,000 K).  
Though the origin of LFBOTs remain unknown, some are thought to originate in the core collapse of stars. 
In any case, their extreme properties make them promising multi-messenger sources in their own right \cite{fang_19,guarini_22}. 

A broad multi-wavelength follow-up campaign was triggered to monitor and characterise SN 2023uqf, including P48-ZTF, P60-Rainbow Camera, LDT-LMI, P200-WASP, P200-WIRC, LCO-Sinistro, Gemini North-GMOS, NOT-ALFOSC, Swift-UVOT, Swift-XRT, Chandra and the VLA (see SM for full details). These observations covered radio, IR, optical, UV and X-ray wavelengths.

The source is detected in the first epoch of UV observations at 11 days post-neutrino, with a host-subtracted u-band magnitude of 19.5 AB mag. This measurement is consistent with thermal emission for a black body of temperature 15,000 K. Substantial cooling is also visible in the optical spectra at this time, with the best-fit thermal continuum falling from 12,500 K at +11d to 9,000 K at +15d.

We fit the optical lightcurve of SN 2023uqf with MOSFiT \citep{mosfit} (Figure \ref{fig:mosfit_lc}, see SM for further details) using the semi-analytical CSM-interaction model of \citet{csm_1}. The CSM fit alone is already able to explain the data relatively well, suggesting that additional Nickel decay emission is not required to power the observed lightcurve. The lightcurve fits provide a tight constraint on the time of supernova explosion (t$_{exp}$ = $-$2.4$^{+1.2}_{-1.3}$d relative to first detection), suggesting that the supernova exploded shortly before the detection of IC231004A. The best fit CSM mass is 0.27$^{+0.41}_{-0.12}$ \Msol, while the best fit ejecta mass is 0.74$^{+2.80}_{-0.56}$ \Msol, both typical for a Ibn (see SM for further details). The CSM mass and ejecta mass here are also broadly consistent with hydrodynamical simulations of Type Ibn supernovae \cite{maeda_22,khatami_24}. However, while these fits are useful to compare our source with other supernovae, we caution that there can be substantial discrepancy between such fits and true physical parameters \citep[see e.g][for more detailed discussion]{moriya_13,moriya_18,suzuki_21}. We emphasize that these parameter estimates include only +/-1$\sigma$ statistical uncertainties and do not include systematic uncertainties from the choice of model.

No X-rays were detected in any of our observations, yielding a deep 3$\sigma$ upper limit of L$_{\textup{x}} < 7 \times 10^{41}$ erg s$^{-1}$ at 25 days after neutrino detection (see SM). We can therefore exclude X-ray emission of the kind observed for AT 2018cow \citep{margutti_19}. Similarly, no radio source was detected in our multi-epoch 10 GHz observations, yielding a 3$\sigma$ upper limit of L$_{\textup{Radio}} < 10^{38}$ erg s$^{-1}$. Taken together, we see no evidence of either a relativistic jet, or of non-thermal emission resembling that of AT 2018cow \citep{margutti_19}. However, our observations cannot constrain X-ray emission from the CSM shock, which has been observed in other interacting supernovae with typical luminosities of order 10$^{40}$ erg s$^{-1}$  or less in \citep{immler_08,dwarkadas_12,pellegrino_24}. We also cannot exclude additional absorption in the source.

To sum up our observations, SN 2023uqf has an extreme lightcurve that is faster and more luminous than most Type Ibn supernovae. However, as illustrated in Figure \ref{fig:Spectrum}, SN 2023uqf clearly resembles a typical Type Ibn supernova spectroscopically. It does not resemble AT 2018cow, the only LFBOT for which a spectroscopic sequence was obtained. The X-ray and radio observations further disfavour non-thermal emission of the kind observed in AT 2018cow and some of the other LFBOTs. SN 2023uqf is thus among the more luminous and rapid objects in the Type Ibn population, and reconfirms that optical lightcurves alone cannot be used to cleanly separate LFBOTs from extreme Type Ibn supernovae \citep[see e.g.][]{shivvers_16,fox_2019,ho_fbot_sample_23,Pellegrino2022,maeda_22}. Whether these two populations form a broader continuum is an open question \citep{fox_2019,xiang_21,metzger_22,Pellegrino2022}. In any case, the spectroscopic observations of SN 2023uqf confirm that this supernova is capable of neutrino production via CSM interaction, and the physical properties of the supernova are also consistent with this picture.

\subsection{Neutrino Production in SN 2023uqf}

We consider in turn the three classes of supernova neutrino production models from the literature for SN 2023uqf:

\begin{itemize}
    \item \textbf{On-Axis Relativistic Jet} - Neutrinos would be expected shortly after supernova explosion \cite{meszaros_01, soebur_05, horiuchi_08,  murase_13, senno_16, ning_18}, alongside non-thermal emission at gamma-ray, X-ray and radio wavelengths. However, no GRB was detected by Fermi-GBM \citep{fermi_gbm} in coincidence with IC231004A \citep{gcn_gbm}, nor has any Type Ibn supernova been previously associated with a long GRB (see \citet{grb_review_21} for a recent review of GRB-supernovae). Our deep and cadenced observations in X-ray and radio further disfavour the presence of any on-axis relativistic jet, as does the lack of a GRB afterglow detection in serendipitous optical observations by ATLAS \citep{atlas_18, atlas_transients_20} at the time of neutrino detection (see also SM).
    \item \textbf{CSM Interaction} - Neutrino production \cite{murase_11,zirakashvili_16,petropolou_17,murase_18,murase_19,wang_19,sarmah_22,murase_24,waxman_24} would be expected to begin after shock breakout or with a delay of a few days after the shock ceases to be radiation mediated \citep{murase_24}, and generally be most intense when the source is optically bright. However, we emphasise that the expected emission time window is broad, and neutrinos would not be expected cluster narrowly around the optical lightcurve peak. Both the timing of the IC231004A, and the clear detection of CSM signatures in SN 2023uqf, indicate that this scenario is viable. 
    \item \textbf{Choked Jet} - Neutrinos would also be expected shortly after explosion \citep{meszaros_01, murase_13, nakar_15, senno_16}, but with no simultaneous electromagnetic signatures to reveal the presence of the jet. Our observations of SN 2023uqf do suggest that there is enough CSM to choke a jet (as little as 10$^{-2}$ \Msol may be sufficient, see e.g \citet{nakar_15}, though other estimates suggest only a weak jet would be choked with such CSM \citep{duffell_20}). Even if the jet is not fully choked, CSM may still cause a relativistic jet to fall below the threshold for detection by GRB monitors. Beyond the CSM itself, the jet can be choked by the outer layers of the star. Type Ibn supernovae are generally believed to arise from massive progenitor stars \citep[see e.g.][]{maeda_22}, with massive progenitors being favoured to launch relativistic jets \cite[see e.g][]{macfayden_99}, and providing additional material to choke a jet. The spectral confirmation of SN2023uqf as an Type Ibn supernova therefore enhances the plausibility of a choked-jet scenario. 
    
    Furthermore, our lightcurve analysis is consistent with neutrino detection close to core collapse. The choked jet scenario is therefore also consistent with our observations, and remains a viable mechanism. High ejecta velocities would be expected in such a scenario, for which we have no evidence in our post-peak spectra, but no prompt data is available. In any case, the velocities observed in optical spectra do not necessarily measure the ejecta velocity itself. For example, \citet{pastorello_16} suggested that the \ion{He}{I} velocities in Type Ibn supernova spectra may arise from the gas interface between the forward and reverse shock, leaving the ejecta velocity unconstrained \citep[see also][]{karamehmetoglu2019snibn}.
\end{itemize} 

To summarise, we conclude that the on-axis relativistic jet model is unlikely, but the presence of CSM interaction mean that both a classical CSM-shock model or a choked jet model for neutrino production would be consistent with our observations.

Based on the latest results of the ZTF Bright Transient Survey \cite{fremling20bts, perley2020bts}, Type Ibn supernovae represent 0.5\% of all supernovae, 2.1\% of core-collapse supernovae and 15.7\% of interacting supernovae in a flux-limited survey. After accounting for all 43 neutrino follow-up campaigns conducted to date, and using the ZTF detection window of 8 days, we estimate that the probability discovering a Type Ibn supernova like SN 2023uqf by chance with our ZTF program is p$=0.28\% \pm 0.06$ (see SM for further details). A chance coincidence is thus disfavoured at the level of $\sim3\sigma$, suggesting that SN 2023uqf is the likely origin of IC231004A, and comparable to other associations reported in the literature \citep[see e.g][]{ic_ps_10_yr,txs_mm,bran,tywin,padovani_16}.

We note that there is an enormous `Eddington Bias' for the detection of individual neutrinos from a source population \citep{nora_eddington}, and an observed coincidence does not imply that the identified counterpart is `special'. Any neutrino source population will have an aggregate flux distributed across many individual objects. Even when these sources are individually dim, with fluxes far below IceCube's detection threshold, the cumulative flux from many dim sources can still be sufficient to expect observed neutrino-source coincidences. Moreover, the combined neutrino flux from sources at higher redshift will typically exceed the emission from very nearby sources (see SM for a more detailed discussion). Our observations should not therefore be interpreted as suggesting that SN 2023uqf is a unique neutrino-bright source, but rather as evidence of aggregate neutrino emission from the broader Ibn population to which SN 2023uqf belongs. 

A single neutrino-Ibn association from the ZTF follow-up program implies that $>$ 1.6\% of high-energy astrophysical neutrino alerts should arise from Type Ibn supernovae, while the lack of additional neutrino-Ibn associations in our ZTF program mean that this population cannot explain all astrophysical neutrino alerts ($<$ 98\%, see SM for more details). These limits are consistent with previous IceCube limits on neutrinos from core-collapse supernovae \citep{ic_sn_23}, and represents the first evidence to support CCSN explosions as hadronic accelerator sites. 


The discovery of SN 2023uqf leaves several questions unanswered. From a single source association, it is not possible to determine the mechanism for neutrino production in this case, nor to draw any broader conclusions about the prevalence of neutrino emission in interacting supernovae. In particular, more associations would be required to understand whether Type Ibn supernovae are uniquely neutrino-bright, or whether more common Type IIn supernova are also important neutrino sources. Given the physical differences between the two populations, it is certainly plausible that they have different neutrino emission properties. 

Finding more SN 2023uqf-like sources will require continued vigilance. Given the fast-fading nature of the source, with a potential classification window of only a few days from first detection, it is unlikely that this supernova would have been recovered serendipitously by ZTF. In particular, almost all detections from Figure \ref{fig:mosfit_lc} occurred in either deep ZTF ToO observations or triggered follow-up with larger telescopes (see SM for further details). Archival neutrino analysis in the style of \citet{ic_sn_23} would have missed this source entirely. SN 2023uqf thus highlights the importance of prompt and systematic neutrino follow-up to reveal neutrino sources, exemplified by the ZTF program. In future, a similar program with the more sensitive \textit{Vera C. Rubin Observatory} \citep{lsst,rubin_too,PSTN-056} will be capable of discovering far more such neutrino-supernova associations \citep{stein_23}, and provide us with the opportunity to comprehensively study multi-messenger supernovae. 

However, even with this discovery of SN 2023uqf, we already have yet more evidence that the astrophysical neutrino flux has diverse origins. Multi-messenger astronomy is continuing to reveal a a rich tapestry of different populations that all contribute to the overall diffuse neutrino flux, including blazars, tidal disruption events, Seyfert galaxies and now supernovae.


\input{acknowledgment_list}

\bibliographystyle{apsrev}
\bibliography{apssamp}

\clearpage

\appendix
\input{base_sm.tex}

\end{document}

%% file: authors.tex
\author{Robert Stein \orcidlink{0000-0003-2434-0387}}
\email{rdstein@umd.edu}
\affiliation{Division of Physics, Mathematics, and Astronomy, California Institute of Technology, 1200 East California Blvd, MC 249-17, Pasadena, CA 91125, USA}
\affiliation{Department of Astronomy, University of Maryland, College Park, MD 20742, USA}
\affiliation{Joint Space-Science Institute, University of Maryland, College Park, MD 20742, USA} 
\affiliation{Astrophysics Science Division, NASA Goddard Space Flight Center, Mail Code 661, Greenbelt, MD 20771, USA} 

\author{Anna Y. Q. Ho \orcidlink{0000-0002-9017-3567}}
\affiliation{Department of Astronomy, Cornell University, Ithaca, NY 14853, USA}

\author{Anjasha Gangopadhyay \orcidlink{0000-0002-3884-5637}}
\affiliation{Oskar Klein Centre, Department of Astronomy, Stockholm University, AlbaNova, SE-106 91 Stockholm, Sweden}

\author{Tomás Ahumada \orcidlink{0000-0002-2184-6430}}
\affiliation{Division of Physics, Mathematics, and Astronomy, California Institute of Technology, 1200 East California Blvd, MC 249-17, Pasadena, CA 91125, USA}

\author{Mansi M. Kasliwal \orcidlink{0000-0002-5619-4938}}
\affiliation{Division of Physics, Mathematics, and Astronomy, California Institute of Technology, 1200 East California Blvd, MC 249-17, Pasadena, CA 91125, USA}

\author{Jannis Necker \orcidlink{0000-0003-0280-7484}}
\affiliation{Deutsches Elektronen-Synchrotron DESY, Platanenallee 6, D-15738 Zeuthen, Germany}
\affiliation{Institut f\"ur Physik, Humboldt-Universit\"at zu Berlin, D-12489 Berlin, Germany}

\author{Simeon Reusch \orcidlink{0000-0002-7788-628X}}
\affiliation{Deutsches Elektronen-Synchrotron DESY, Platanenallee 6, D-15738 Zeuthen, Germany}
\affiliation{Institut f\"ur Physik, Humboldt-Universit\"at zu Berlin, D-12489 Berlin, Germany}

\author{Marek Kowalski \orcidlink{0000-0001-8594-8666}}
\affiliation{Deutsches Elektronen-Synchrotron DESY, Platanenallee 6, D-15738 Zeuthen, Germany}
\affiliation{Institut f\"ur Physik, Humboldt-Universit\"at zu Berlin, D-12489 Berlin, Germany}

\author{Anna Franckowiak \orcidlink{0000-0001-8594-8666}}
\affiliation{Ruhr University Bochum, Faculty of Physics and Astronomy, Astronomical Institute (AIRUB), Universitätsstraße 150, 44801 Bochum, Germany}

\author{Jesper Sollerman \orcidlink{0000-0003-1546-6615}}
\affiliation{Oskar Klein Centre, Department of Astronomy, Stockholm University, AlbaNova, SE-106 91 Stockholm, Sweden}

\author{Kohta Murase \orcidlink{0000-0002-5358-5642}}
\affiliation{Department of Physics, The Pennsylvania State University, University Park, PA 16802, USA}
\affiliation{Department of Astronomy \& Astrophysics, The Pennsylvania State University, University Park, PA 16802, USA}
\affiliation{Center for Multimessenger Astrophysics, Institute for Gravitation and the Cosmos, The Pennsylvania State University, University Park, PA 16802, USA}
\affiliation{Center for Gravitational Physics and Quantum Information, Yukawa Institute for Theoretical Physics, Kyoto University, Kyoto, Kyoto 606-8502, Japan}


\author{Igor Andreoni \orcidlink{0000-0002-8977-1498}}
\affiliation{Department of Astronomy, University of Maryland, College Park, MD 20742, USA}
\affiliation{Joint Space-Science Institute, University of Maryland, College Park, MD 20742, USA} 
\affiliation{Astrophysics Science Division, NASA Goddard Space Flight Center, Mail Code 661, Greenbelt, MD 20771, USA} 
\affiliation{Department of Physics and Astronomy, University of North Carolina at Chapel Hill, Chapel Hill, NC 27599-3255, USA}

\author{Eric C. Bellm \orcidlink{0000-0001-8018-5348}}
\affiliation{DIRAC Institute, Department of Astronomy, University of Washington, 3910 15th Avenue NE, Seattle, WA 98195, USA}

\author{Joshua Bloom \orcidlink{0000-0002-7777-216X}}
\affiliation{Department of Astronomy, University of California, Berkeley, CA 94720}

\author{Se{\'a}n J. Brennan \orcidlink{0000-0003-1325-6235}}
\affiliation{Oskar Klein Centre, Department of Astronomy, Stockholm University, AlbaNova, SE-106 91 Stockholm, Sweden}

\author{Liam Connor \orcidlink{0000-0002-7587-6352}}
\affiliation{Center for Astrophysics | Harvard \& Smithsonian, Cambridge, MA 02138-1516, USA}

\author{Michael W. Coughlin \orcidlink{0000-0002-8262-2924}}
\affiliation{School of Physics and Astronomy, University of Minnesota, Minneapolis, Minnesota 55455, USA}

\author{Richard Dekany \orcidlink{0000-0002-5884-7867}}
\affiliation{Caltech Optical Observatories, California Institute of Technology, Pasadena, CA 91125, USA}

\author{Andrew Drake}
\affiliation{Division of Physics, Mathematics, and Astronomy, California Institute of Technology, 1200 East California Blvd, MC 249-17, Pasadena, CA 91125, USA}

\author{Christoffer Fremling \orcidlink{0000-0002-4223-103X}}
\affiliation{Caltech Optical Observatories, California Institute of Technology, Pasadena, CA 91125, USA}
\affiliation{Division of Physics, Mathematics, and Astronomy, California Institute of Technology, 1200 East California Blvd, MC 249-17, Pasadena, CA 91125, USA}

\author{Ariel Goobar \orcidlink{0000-0002-4163-4996}}
\affiliation{Oskar Klein Centre, Department of Physics, Stockholm University, AlbaNova, SE-106 91 Stockholm, Sweden}

\author{Matthew J. Graham \orcidlink{0000-0002-3168-0139}}
\affiliation{Division of Physics, Mathematics, and Astronomy, California Institute of Technology, 1200 East California Blvd, MC 249-17, Pasadena, CA 91125, USA}

\author{Steven L. Groom \orcidlink{0000-0001-5668-3507}}
\affiliation{IPAC, California Institute of Technology, 1200 E. California Blvd, Pasadena, CA 91125, USA}

\author{Theophile Jegou du Laz \orcidlink{0009-0003-6181-4526}}
\affiliation{Division of Physics, Mathematics, and Astronomy, California Institute of Technology, 1200 East California Blvd, MC 249-17, Pasadena, CA 91125, USA}

\author{Daniel Perley \orcidlink{0000-0001-8472-1996}}
\affiliation{Astrophysics Research Institute, Liverpool John Moores University, 146 Brownlow Hill, Liverpool L3 5RF, UK}

\author{Priscila J. Pessi \orcidlink{0000-0002-8041-8559}}
\affiliation{Oskar Klein Centre, Department of Astronomy, Stockholm University, AlbaNova, SE-106 91 Stockholm, Sweden}

\author{Josiah Purdum \orcidlink{0000-0003-1227-3738}}
\affiliation{Caltech Optical Observatories, California Institute of Technology, Pasadena, CA 91125, USA}

\author{Brendan O'Connor \orcidlink{0000-0002-9700-0036}}
\affiliation{McWilliams Center for Cosmology and Astrophysics, Department of Physics, Carnegie Mellon University, Pittsburgh, PA 15213, USA}

\author{Steve Schulze \orcidlink{0000-0001-6797-1889}}
\affiliation{Center for Interdisciplinary Exploration and Research in Astrophysics (CIERA), Northwestern University, 1800 Sherman Ave, Evanston, IL 60201, USA}

\author{Gokul P. Srinivasaragavan \orcidlink{0000-0002-6428-2700}}
\affiliation{Department of Astronomy, University of Maryland, College Park, MD 20742, USA}
\affiliation{Joint Space-Science Institute, University of Maryland, College Park, MD 20742, USA} 
\affiliation{Astrophysics Science Division, NASA Goddard Space Flight Center, Mail Code 661, Greenbelt, MD 20771, USA} 

\author{Sylvain Veilleux \orcidlink{0000-0002-3158-6820}}
\affiliation{Department of Astronomy, University of Maryland, College Park, MD 20742, USA}
\affiliation{Joint Space-Science Institute, University of Maryland, College Park, MD 20742, USA} 

\author{Avery Wold \orcidlink{0000-0002-9998-6732}}
\affiliation{IPAC, California Institute of Technology, 1200 E. California Blvd, Pasadena, CA 91125, USA}

\author{Lin Yan \orcidlink{0000-0003-1710-9339}}
\affiliation{Caltech Optical Observatories, California Institute of Technology, Pasadena, CA 91125, USA}

%% file: acknowledgment_list.tex
\subsection{Acknowledgments}
\begin{acknowledgments}
R.S. would like to acknowledge and thank Avishay Gal-Yam for fruitful discussions on neutrinos from supernovae.

R.S. and M.M.K acknowledge support from grants by the National Science Foundation  (AST 2206730) and the David and Lucille Packard Foundation (PI Kasliwal).
A.Y.Q.H. acknowledges support from NASA Grant 80NSSC23K1155. 
J.N. was supported by the Helmholtz Weizmann Research School on Multimessenger Astronomy, funded through the Initiative and Networking Fund of the Helmholtz Association, DESY, the Weizmann Institute, the Humboldt University of Berlin, and the University of Potsdam.
A.F. acknowledges support from the DFG via the Collaborative Research Center SFB1491 Cosmic Interacting Matters – From Source to Signal.
J.S.B. acknowledges support from the Gordon and Betty Moore Foundation, through both the Data-Driven Investigator Program and a dedicated grant, provided critical funding for SkyPortal.
M.W.C. acknowledges support from the National Science Foundation with grant numbers PHY-2308862, PHY-2117997 and PHY-2409481.
B. O. is supported by the McWilliams Postdoctoral Fellowship at Carnegie Mellon University.
S. S. is partially supported by LBNL Subcontract 7707915.

Based on observations obtained with the Samuel Oschin Telescope 48-inch and the 60-inch Telescope at the Palomar Observatory as part of the Zwicky Transient Facility project. ZTF is supported by the National Science Foundation under Grants No. AST-1440341 and AST-2034437 and a collaboration including current partners Caltech, IPAC, the Weizmann Institute of Science, the Oskar Klein Center at Stockholm University, the University of Maryland, Deutsches Elektronen-Synchrotron and Humboldt University, the TANGO Consortium of Taiwan, the University of Wisconsin at Milwaukee, Trinity College Dublin, Lawrence Livermore National Laboratories, IN2P3, University of Warwick, Ruhr University Bochum, Northwestern University and former partners the University of Washington, Los Alamos National Laboratories, and Lawrence Berkeley National Laboratories. Operations are conducted by COO, IPAC, and UW. 

SED Machine is based upon work supported by the National Science Foundation under Grant No. 1106171.

The Gordon and Betty Moore Foundation, through both the Data-Driven Investigator Program and a dedicated grant, provided critical funding for SkyPortal. 

The authors wish to recognize and acknowledge the very significant cultural role and reverence that the summit of Maunakea has always had within the Native Hawaiian community. We are most fortunate to have the opportunity to conduct observations from this mountain.

Some of the data presented herein were obtained at Keck Observatory, which is a private 501(c)3 non-profit organization operated as a scientific partnership among the California Institute of Technology, the University of California, and the National Aeronautics and Space Administration. The Observatory was made possible by the generous financial support of the W. M. Keck Foundation. 

Based on observations obtained at the international Gemini Observatory, a program of NSF NOIRLab, which is managed by the Association of Universities for Research in Astronomy (AURA) under a cooperative agreement with the U.S. National Science Foundation on behalf of the Gemini Observatory partnership: the U.S. National Science Foundation (United States), National Research Council (Canada), Agencia Nacional de Investigaci\'{o}n y Desarrollo (Chile), Ministerio de Ciencia, Tecnolog\'{i}a e Innovaci\'{o}n (Argentina), Minist\'{e}rio da Ci\^{e}ncia, Tecnologia, Inova\c{c}\~{o}es e Comunica\c{c}\~{o}es (Brazil), and Korea Astronomy and Space Science Institute (Republic of Korea).

Based on observations made with the Nordic Optical Telescope, owned in collaboration by the University of Turku and Aarhus University, and operated jointly by Aarhus University, the University of Turku and the University of Oslo, representing Denmark, Finland and Norway, the University of Iceland and Stockholm University at the Observatorio del Roque de los Muchachos, La Palma, Spain, of the Instituto de Astrofisica de Canarias.
Horizon 2020/2021-2025: This project has received funding from the European Union’s Horizon 2020 research and innovation programme under grant agreement No 101004719 (ORP: OPTICON RadioNet Pilot)
The data presented here were obtained in part with ALFOSC, which is provided by the Instituto de Astrofisica de Andalucia (IAA) under a joint agreement with the University of Copenhagen and NOT.

These results made use of Lowell Observatory’s Lowell Discovery Telescope (LDT), formerly the Discovery Channel Telescope. Lowell operates the LDT in partnership with Boston University, Northern Arizona University, the University of Maryland, and the University of Toledo. Partial support of the LDT was provided by Discovery Communications. LMI was built by Lowell Observatory using funds from the National Science Foundation (AST-1005313).

This paper is based on observations made with the MuSCAT instruments, developed by the Astrobiology Center (ABC) in Japan, the University of Tokyo, and Las Cumbres Observatory (LCOGT). MuSCAT3 was developed with financial support by JSPS KAKENHI (JP18H05439) and JST PRESTO (JPMJPR1775), and is located at the Faulkes Telescope North on Maui, HI (USA), operated by LCOGT. MuSCAT4 was developed with financial support provided by the Heising-Simons Foundation (grant 2022-3611), JST grant number JPMJCR1761, and the ABC in Japan, and is located at the Faulkes Telescope South at Siding Spring Observatory (Australia), operated by LCOGT.


The scientific results reported in this article are based on observations made by the Chandra X-ray Observatory.

The National Radio Astronomy Observatory is a facility of the National Science Foundation operated under cooperative agreement by Associated Universities, Inc.
\end{acknowledgments}

%% file: base_sm.tex
\renewcommand{\figurename}{SM FIG}
\renewcommand{\tablename}{SM TABLE}
\setcounter{figure}{0}   

\section{Supplementary Material}

\section{ZTF Candidate Selection}

We briefly describe the procedure by which candidate counterparts are selected by the ZTF Neutrino Follow-Up Program. Further details of this procedure are given in \citet{bran} and \citet{stein_23}. 

All ZTF raw images are processed by IPAC \citep{masci_19}, generating calibrated science images which are then subtracted against dedicated reference images to produce a `difference' image. Source detection is performed on these difference images, yielding a set of detections of transient/variable sources in the image. Each detections is packaged as an \texttt{Avro} alert \citep{zads}, including additional contextual information such as previous detections, results from cross-matching to external catalogues and machine-learning classifier scores. Alerts are distributed via Kafka streams to ZTF alert brokers.

The ZTF neutrino program analyses ZTF data with the dedicated \texttt{nuztf} analysis code \citep{nuztf}, which queries the \texttt{AMPEL} alert broker archive to retrieve ZTF data \citep{ampel_19}. Alerts are selected if they lie within a given neutrino spatial localisation are selected, and if they were detected in a 14 day window after the neutrino arrival time. 

These ZTF alerts are loaded and then filtered by \texttt{nuztf}, to select probable extra-galactic transients or variables. The following cuts are applied:

\begin{itemize}
	\item \textbf{Image artefact rejection}: ZTF alerts contain the score from a deep-learning classifier trained to distinguish `real' detections from `bogus ones' \citep{ztf_drb}. We select alerts with a `deep real/bogus' (drb) score greater than 0.3, for which a false negative rate of $<$0.2\% is expected \citep{ztf_drb}.
	\item \textbf{Solar System Objects:} We reject alerts which have been cross-matched to known solar system objects. We further require at least two detections of an object separated by at least 15 minutes, which removes moving objects even if they have not been catalogued.
	\item \textbf{Stars:} We remove known sources which have been classified in PS1 data \citep{ps1} as likely stars using morphology \citep{sgscore}. We further flag sources with high ($>$3$\sigma$) measured parallax in Gaia DR3 \citep{gaia_dr3}. 
\end{itemize}

The combination of drb cuts and our requirement for multiple detections mean that the overwhelming majority of sources selected by \texttt{nuztf} are ultimately real astrophysical sources. We subsequently vet remaining candidates by visual inspection, removing stars, remaining artifacts and any AGN which are not flaring. These steps are similar to those employed widely in the community, for example the ZTF Bright Transient Survey \citep{fremling20bts,perley2020bts}.

Applying this procedure for IC231004A, we find 4 sources for vetting (including SN 2023uqf). Of these sources, one is a star, one is a lone faint detection of a known AGN in Milliquas without any evidence of flaring \citep{milliquas}, and one is a faint nuclear source with no evolution over several thousand days that is likely a subtraction artefact. None of the three sources is consistent with being a transient or flaring neutrino source, leaving SN 2023uqf as the only potential counterpart in ZTF data \citep{gcn_disc}. 
This single candidate is consistent with previous follow-up campaigns in our program \citep{stein_23}, with candidate counterparts selected at a rate of $\sim$1 per 3 sq degrees.

\section{Photometric Observations}

Photometric monitoring of SN 2023uqf in the optical was performed with:

\begin{itemize}
	\item the Rainbow Camera of the 1.5m Telescope (P60, \citep{blagorodnova_18}) at Palomar Observatory (PI: Franckowiak), with reduction using the standard \texttt{FPipe} imaging pipeline \citep{fpipe}.
	\item the Large Monolithic Imager (LMI) of the 4.3m Lowell Discovery Telescope (LDT, PI: Andreoni), with reductions performed using a custom image reduction and photometry pipeline.
	\item the Wafer-Scale camera for Prime (WASP) at the 5.1m Palomar Hale Telescope (P200, PI: Kulkarni). Reductions were performed with the WASP pipeline implemented in \texttt{mirar} \citep{mirar}. 
	\item the Alhambra Faint Object Spectrograph and Camera (ALFOSC) at the 2.56m Nordic Optical Telescope (NOT, PI: Sollerman). Reductions were performed with the \texttt{AUTOPHOT} pipeline \citep{autophot_22}. 
	\item the Multicolor Simultaneous Camera for studying Atmospheres of Transiting exoplanets (MUSCAT, \citep{muscat_3}) on the 2m telescope operated by Las Cumbres Observatory \citep{lco_13} (PI: Das). Data were reduced with a custom python imaging pipeline.
	\item the Gemini Multi-Object Spectrograph North (GMOS-N, \citep{gmosn}) at the 8.1m Gemini North telescope (PI: Stein). Initial detrending were reduced with the \texttt{DRAGONS} pipeline \citep{dragons} and subsequent data reduction was performed with in \texttt{mirar} \citep{mirar}.
	\item the Wide Field Infrared Camera (WIRC, PI: Connor and Kasliwal) at the 5.1m Palomar Hale Telescope. Reductions were performed with the WIRC pipeline implemented in \texttt{mirar} \citep{mirar}.
	\item the Ultra-Violet Optical Telescope (UVOT, \citep{swift_uvot}) on board the \textit{Neil Gehrels Swift Observatory} \citep{gehrels_04}. Data were reduced using \texttt{HEASoft} \citep{heasoft}. 
\end{itemize}

Image subtraction for optical filters ($g$, $r$, $i$ $z$) was performed using either \texttt{ZOGY} \citep{zogy} or \texttt{HOTPANTS} \citep{hotpants} against reference images from Pan-STARRS1 \citep{ps1}. For WIRC, image subtraction was performed using ZOGY against reference images from UKIRT \citep{ukirt}. No image subtraction was performed for UV images. However, the host was detected in SDSS archival imaging, with a u-band magnitude of 22.87, suggesting the vast majority of the measured UV flux belongs to the transient.

All observations of SN 2023uqf are listed in Table \ref{tab:photometry}.

\section{SN 2023uqf in Duration/Luminosity Space}
\begin{figure}
	\centering
	\includegraphics[width=\linewidth]{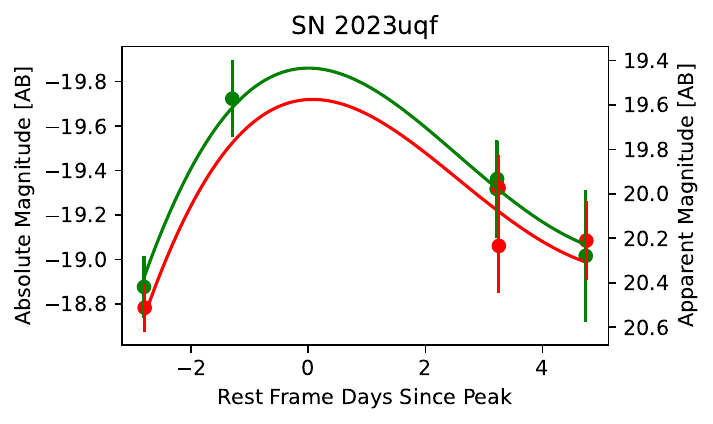}
	\includegraphics[width=\linewidth]{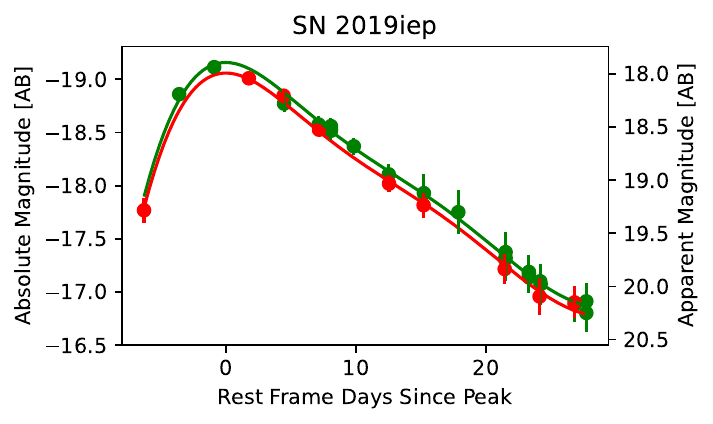}
	\caption{Demonstration of the Gaussian Process fitting procedure on ZTF data for SN 2023uqf (top) and an example source from the ZTF Type Ibn  sample (SN 2019iep, bottom). The maximum of the $g-$band fit is taken as the peak.}
	\label{fig:gaussian_process}
\end{figure}

We analyse the ZTF lightcurve of SN 2023uqf using a simple Gaussian Process two-band fitting procedure originally developed for Tidal Disruption Events (see \citet{tdescore} for more details). Data was fit in log(phase + 20d) space following the procedure introduced in \citet{Hosseinzadeh2017}. The lightcurve is modelled iteratively assuming that the g-r colour of each source evolves linearly over time, an approximation which holds well for Ibn supernovae around peak. The best fit is given in SM Figure \ref{fig:gaussian_process}. We then extract t$_{1/2}$, defined as the time in rest frame days that the source spends above half the $g-$band peak flux.

We repeat the same procedure on a sample of 34 ZTF Type Ibn supernovae. These Type Ibn supernovae were all selected systematically from the Zwicky Transient Factory archive including the Bright Transient Survey sample explorer applying an apparent magnitude cut of 19.5 and Milky way extinction cut of less than 0.3 mag. The sample also includes the Type Icn supernovae which were misclassified early on and re-classified as a Type Ibn supernova. The detailed analysis of this sample will be presented in \citet{Gangopadhyay2025}.

\begin{figure}
	\centering
	\includegraphics[width=0.45\textwidth]{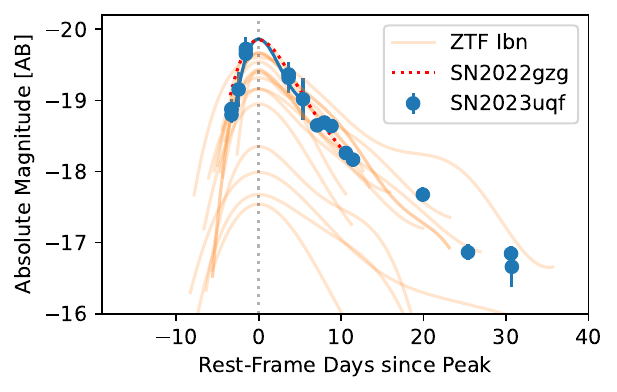}
	\caption{The full $g$-band lightcurve of SN 2023uqf (blue scatter points), and the Gaussian Process fit to the ZTF data of the SN 2023uqf (solid blue line). For comparison, similar lightcurve fits for every supernova in the ZTF Ibn sample are shown in orange. One other Type Ibn (SN 2022gzg, marked with the red dashed line) has a similar luminosity to SN 2023uqf, but with slower evolution.}
	\label{fig:rise_vs_peak}
\end{figure}

An example Type Ibn fit is shown by SM Figure \ref{fig:gaussian_process}. Of these, we remove two sources which have a double peak and a further 17 which either do not have a resolved rise or resolved fade in the light curve. This leaves us with 17 sources with a measured duration, plotted in main text Figure 2. We also plot lightcurve models for SN 2023uqf and each member of the ZTF Type Ibn sample in SM Figure \ref{fig:rise_vs_peak}. We consider only ZTF photometry for these supernovae and for SN 2023uqf, to avoid fitting bias from varying degrees of additional follow-up conducted. We similarly omit detections arising from manual requests to the ZTF forced photometry service operated by IPAC \citep{masci_23}, to provide a fair comparison between SN 2023uqf and the other Type Ibn supernovae. 

\begin{figure}
	\centering
	\includegraphics[width=0.45\textwidth]{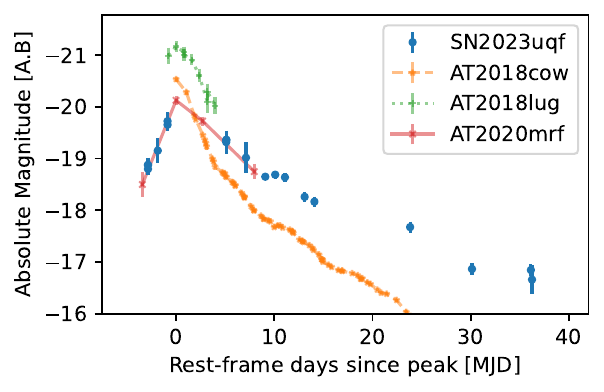}
	\caption{Observed $g$-band lightcurve of SN 2023uqf (blue) assuming that the peak date occurred one day after the brightest ZTF detection, alongside some LFBOTs from the literature \cite{perley_at2018cow,at2018lug,yao_22_mrf}. SN 2023uqf matches particularly well to AT 2020mrf (red) \citep{yao_22_mrf}. 
	}
	\label{fig:Lightcurve}
\end{figure}

We also present a comparison of the lightcurve of SN 2023uqf to known LFBOTs in SM Figure \ref{fig:Lightcurve}. As was already noted in main text Figure 2, SN 2023uqf resembles the LFBOT AT 2020mrf \citep{yao_22_mrf}.

\section{Lightcurve Analysis}

\begin{figure*}
	\centering
	\includegraphics[width=0.98\textwidth]{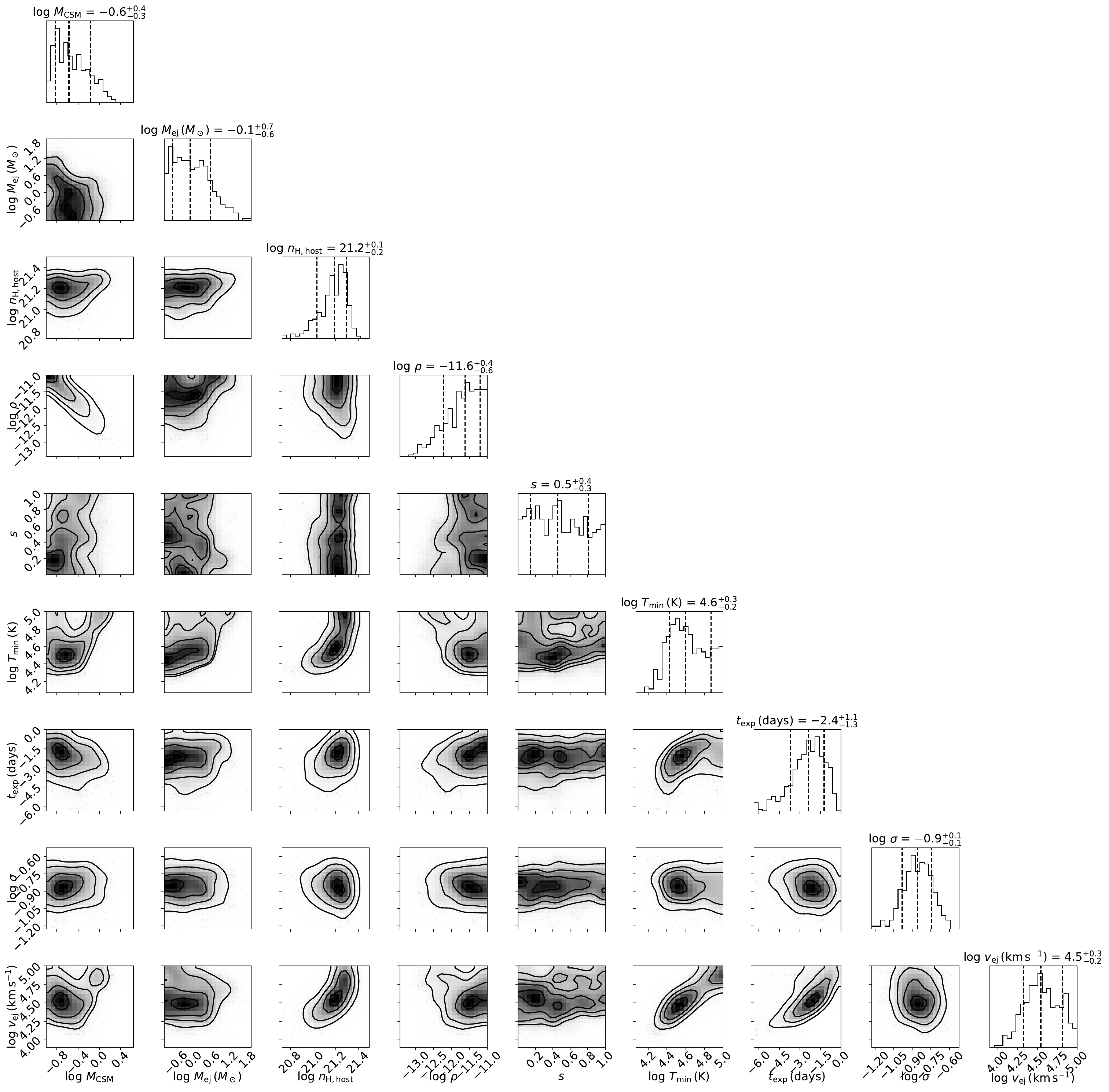}
	\caption{Corner plot of the posteriors from the \mosfit lightcurve fitting of SN 2023uqf (as shown in main text Figure 2).}
	\label{fig:mosfit_corner}
\end{figure*}

We model the lightcurve using the Modular Open Source Fitter for Transients (\mosfit,  \citep{mosfit}), an open source python code providing semi-analytic lightcurve fitting using a variety of pre-built astrophysical transient models. We use the \texttt{CSM} supernova model 
\citep{csm_1,csm_2,csm_3}, which includes contributions only from CSM interaction without any additional $^{56}$Ni decay. This is a nearly valid approximation as most SNe~Ibn/Icns show a negligible fraction of $^{56}$Ni contributing to the lightcurve of the SNe. The model assumes a single zone, with CSM following a power law distribution, where the kinetic energy of the forward shock and reverse shock powers the lightcurve \citep{csm_3}. We run with 400 walkers and 20000 iterations. We restrict our fit to those detections which are high signal-to-noise, defined as those at least one sigma above the image limiting magnitude. This removes some marginal detections from the fit. A full list of priors and fit values is given in Table \ref{tab:mosfit_parameters}. The best-fit results are shown in main text Figure 3, and the resulting corner plot is shown in SM Figure \ref{fig:mosfit_corner}.

\begin{table}[]
	\centering
	\begin{tabular}{c|c|c|c}
		Parameter & Lower & Upper & Best \\
		&Bound& Bound & Fit \\
		\hline
		$\log\, M_{\rm CSM}$ & -1.0 & 2.0 & $-0.57^{+0.40}_{-0.25}$ \\
		$\log\, M_{\rm ej}\,(M_\odot)$ & -1.0 & 2.0 & $-0.13^{+0.68}_{-0.60}$ \\
		$\log\, n_{\rm H,host}$ & 16.0 & 23.0 & $21.19^{+0.10}_{-0.15}$ \\
		$\log\, \rho$ & -15.0 & -11.0 & $-11.61^{+0.42}_{-0.60}$ \\
		$s$ & 0.0 & 1.0 & $0.46^{+0.35}_{-0.31}$ \\
		$\log\, T_{\min}\,{\rm (K)}$ & 0.0 & 5.0 & $4.60^{+0.27}_{-0.18}$ \\
		$t_{\rm exp}\,{\rm (days)}$ & -50.0 & 0.0 & $-2.38^{+1.15}_{-1.33}$ \\
		$\log\, \sigma$ & -5.0 & 1.0 & $-0.86^{+0.11}_{-0.12}$ \\
		$\log\, v_{\rm ej}\,({\rm km\,s}^{-1})$ & 3.0 & 5.0 & $4.54^{+0.27}_{-0.21}$ \\
	\end{tabular}
	\caption{Priors and best-fit values for the ten parameters of the \mosfit \texttt{CSM} model. Uniform priors were used throughout. $t_{\rm exp}$ is defined relative to the first ZTF detection (i.e +0.9d from neutrino detection).}
	\label{tab:mosfit_parameters}
\end{table}

\citet{Pellegrino2022} inferred the properties of a sample of SNe~Ibn through bolometric lightcurve fitting, providing context for our analysis of SN 2023uqf. Typical SNe Type Ibn have CSM masses between 0.2 -- 0.8 \Msol, and ejecta masses between 0.7 -- 1.5 \Msol, consistent with our fit of SN 2023uqf. Generally Type Ibn SNe have lower estimated CSM masses than Type IIn SNe, and the CSM mass and ejecta mass inferred for SN 2023uqf are consistent with the estimates from other Type Ibn SNe.

\section{Spectroscopic Observations}

\begin{figure}
	\centering
	\includegraphics[width=0.45\textwidth]{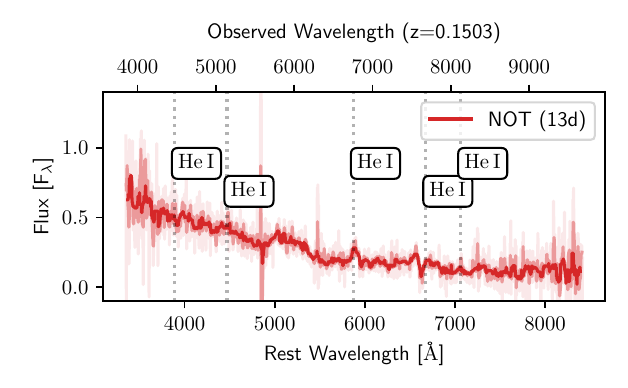}
	\caption{First spectrum of SN 2023uqf, taken with NOT/ALFOSC. He lines are highlighted with vertical dashed lines.}
	\label{fig:spectra_early}
\end{figure}

\begin{figure*}
	\centering
	\includegraphics[width=0.9\textwidth]{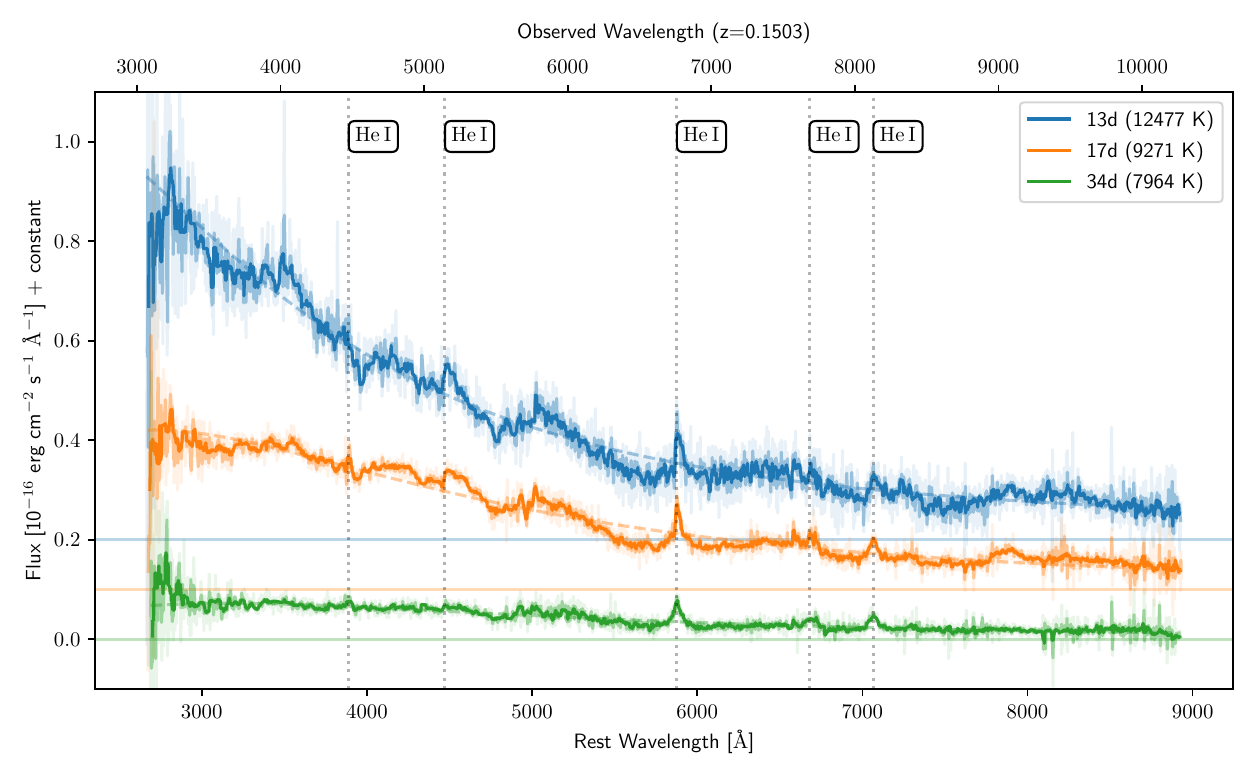}
	\caption{Flux-calibrated evolution of SN 2023uqf in Keck-I/LRIS spectra, relative to time of neutrino detection. The solid line shows the median-smoothed spectrum while the horizontal dotted lines indicates the zero-flux baseline for each spectrum. The dashed lines indicate the best-fit thermal continuum. The vertical dotted lines indicate the He I lines, which are clearly detected in all of our spectra. Prominent zero-velocity galaxy emission lines  (H, O) are masked and interpolated over.}
	\label{fig:spectra_series}
\end{figure*}

\begin{figure}
	\centering
	\includegraphics[width=0.4\textwidth]{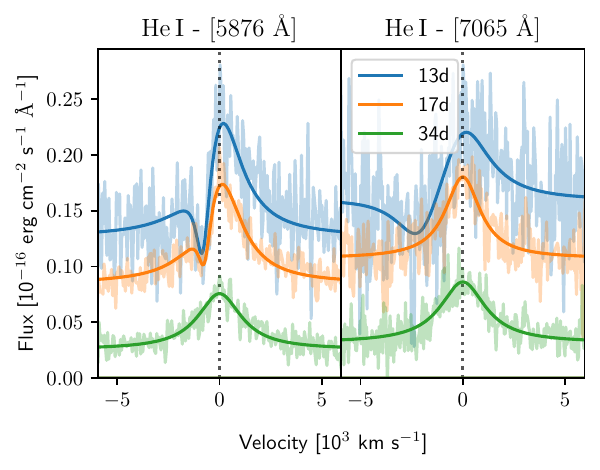}
	\caption{Absolute flux-calibrated spectral evolution of SN 2023uqf emission lines. All spectra were taken with Keck-I LRIS. The \ion{He}{I} lines at 5876 $\rm \AA$ and 7065 $\rm \AA$ are initially asymmetric with a substantial red component, but by the late-time spectrum, they are blueshifted. The P-Cygni profile disappears.  The solid lines show the fitted model.}
	\label{fig:spectra_lines}
\end{figure}

Spectroscopic observations of SN~2023uqf were first obtained with ALFOSC (PI: Franckowiak) on the 2.56m Nordic Optical Telescope (NOT) at 2023-10-17 04:16:15 UT (see SM Figure \ref{fig:spectra_early}), and reduced using the \texttt{PyNOT} pipeline \footnote{\url{https://jkrogager.github.io/pynot/}}. A higher-SNR spectrum was obtained with the Low Resolution Imaging Spectrometer (LRIS \citep{oke_95}) on the 10m Keck-I telescope  shortly afterwards at 2023-10-17 15:29:02 UT (PI: Ravi). Additional LRIS spectra were obtained at 2023-10-21 15:16:07 UT (PI: Kasliwal) and 2023-11-09 13:26:29.212 (PI: Kulkarni). All LRIS spectra were reduced using \texttt{LPipe} \citep{lpipe}.

The spectral evolution of SN 2023uqf from $13$~days to $34$~days post discovery is shown in SM Figure~\ref{fig:spectra_series}. The first LRIS spectrum of SN~2023uqf shows a distinct blue continuum, similar to that of other fast transients \citep{2014Drout,Pellegrino2022}. We fit each LRIS spectrum with a thermal continuum model, and find that the apparent photospheric temperature falls from $\sim$12,500 K at +11d to $\sim$8,000 K in our final spectrum at +32d. However, we note that the SEDs at these temperatures peak in the UV regime \citep{2014Drout} and precise measurements of the temperature would require observations of the UV brightness of the object. 

We see typical features of emission in SN~2023uqf around $\sim$4660 \AA~ in the earliest spectrum, which arises as signature of flash ionised CSM \citep{Gal-Yam2014}. The flash ionization spectral signatures vanished by 17 days. Previously flash ionisation signatures of C III and He II, have been seen in many Type Ibn SNe (PTF12ldy, iPTF15ul \citep{Hosseinzadeh2017}, SN 2010al \citep{Pastorello2015}, SN 2019uo \citep{Gangopadhyay2020} and SN 2019wep \citep{Gangopadhyay2022}). \citet{Galyam2014} had shown that for SN~2013cu, there was an elevated mass loss that produced a wind similar to that seen in WR stars (WR winds). However, \citet{Groh2014} estimated the physical parameters of the Type IIb SN 2013cu and found that the progenitor was likely not a WR star even though the flash ionized features showed WR-like lines. 

In the early spectra, we also see an absorption feature close to $\sim$4000 \AA\ which could have multiple origin. One of the reason for the origin of this feature could be O II \citep{Quimby2011}, but it usually originates in a bunch of absorption lines. It could be due to He II features, originating due to recombination of flash ionized CSM. \citet{Bruch2021} studied a sample of SNe with flash ionisation signatures which had a dip at 4000 \AA, but the origin was said to be not known.  There is a blend of Fe II throughout the evolution of SN~2023uqf between 4000 to 5000 \AA, as is typical for Type Ibn supernovae. So, there is most likely a plethora of Fe lines along with some absorption features whose origin is not well discerned.

As seen clearly in SM Figure \ref{fig:spectra_lines}, we see a distinct He I P-Cygni feature at 17 days that is superimposed on a broader base (the continuum is not flat). This feature is especially prominent at  He I $\lambda5876$, and has been seen in many other Type Ibn supernovae \citep{Hosseinzadeh2017,karamehmetoglu2019snibn}. These features disappear entirely by day 34 as seen in other objects. At +34 d, we see broad emission of He I appearing in the spectral sequence of SN~2023ufq. 

At early phases, the narrow lines are formed in the photosphere located in a dense shell. This shell is either continuously photoionized by early ejecta-CSM interaction in the inner CSM regions and/or by the initial shock breakout. Once recombination occurs, the shell becomes transparent and we see the signatures of underlying SN ejecta and the spectrum is dominated by broad P-Cygni lines \citep{Gangopadhyay2020}. We see the simultaneous presence of a very narrow feature at early phases and transitioning to broad features at +34 d which suggests that these features arise from different emitting regions: the broader He I P-Cygni features are likely a signature of the SN ejecta, while the narrow He I P-Cygni lines are generated in the unperturbed, He-rich CSM. In the earliest spectrum, blueshifted absorption along the line of sight is indicative of the fact that there is an interaction between the ejecta and the CSM towards the observer. 

\citet{Hosseinzadeh2017} have categorised the early narrow features into narrow “P-cygni” and narrow “emission” sub-class depending on whether the CSM shell/blob is viewed edge-on or face-on. This interpretation has been challenged by \citet{karamehmetoglu2019snibn}, who argue that line formation in such SNe is primarily influenced by the physical conditions—namely the density, temperature, and optical depth of the CSM. Through spectral modelling of SN~2018bcc, they demonstrated that the helium features arise from a combination of P-Cygni and pure emission profiles, driven by the high optical depths and densities present in the CSM. 

Even at late times, the helium lines remain optically thick but appear as emission-dominated due to the absence of alternative decay pathways. Ionization and recombination of helium are primarily triggered by high-energy UV and X-ray photons generated near the shock interface. While most emission and electron scattering occur in the ionized layers beyond the shock, P-Cygni absorption features emerge from regions with lower optical depth ($\tau$ $\leq$ 1). 

Penetrating X-rays can fill in these absorption troughs, converting them into emission lines. The red part of the spectrum with He I lines ($\geq$ 6000 \AA) typically show strong emission as they cannot branch into other lines. The blue lines of He I ($\leq$ 6000 \AA) instead trap photons and then these photons emerge at longer wavelengths, resulting in P-Cygni absorptions without the strong emission seen in the redder lines. This provides a compelling alternative explanation for the observed line profiles and likely reflects the phase in which SN~2023ufq changes from P-cygni to emission He features. The spectra at +34 d also shows a blueshifted profile which could possibly be due to dust formation \citep{Nozawa_2006jc}.

\section{X-ray observations of SN 2023uqf}

Target-of-opportunity observations of SN\,2023uqf were obtained with the X-Ray Telescope (XRT \citep{Burrows2005}) on board the Neil Gehrels \emph{Swift} observatory \citep{Gehrels2004} under two programs \footnote{Target ID 16282; TOO ID 19505 PI Pasham, 19539 PI Ho}. XRT data were reduced using the online tool \footnote{\url{https://www.swift.ac.uk/user_objects/}} from the \emph{Swift} team \citep{Evans2007,Evans2009}. 

Observations of SN 2023uqf with the Chandra X-ray Observatory \citep{chandra} were later obtained under a dedicated DDT program (PI: Stein \footnote{\url{https://doi.org/10.25574/29036}}). The exposure was split into two discrete observation windows, beginning Oct 29 2023 9:11AM and Oct 30 2023 6:26AM, with  a total exposure time of 25ks. Observations were conducted using the `S-array' of the Advanced CCD Imaging Spectrometer (ACIS)
\citep{chandra_acis}, with no grating and the target centered on chip S3.

Data were reduced using the `Chandra Interactive Analysis of Observations' (\texttt{CIAO}) software package \citep{chandra_ciao}. No source is seen at the position of SN 2023uqf, yielding an upper limit of $2 \times 10^{-4}$ cts per second at 3$\sigma$ confidence. This corresponds to a 0.3 -- 10.0 keV flux limit of $1.09 \times 10^{-14}$ erg cm$^{-2}$ s$^{-1}$, or $L_{X} < 6.7 \times 10^{41}$ erg s$^{-1}$ at the distance of SN 2023uqf (723 Mpc for a redshift of z=0.1503). All our X-ray data are summarised in Table \ref{tab:xray}. For both instruments, flux limits were derived assuming a power law of index 2.0 and a galactic column density of $2.15 \times 10^{20} \textup{cm}^{2}$.

\begin{table*}[]
	\centering
	\begin{tabular}{c | c | c | c| c | c | c }
		Start Date & Phase & Instrument & Exposure & Count Limit & Flux Limit & Luminosity Limit \\
		(UT) &  [days] & &[ks] & [cts s$^{-1}$] & [10$^{-14}$ erg s$^{-1}$ cm$^{-2}$]&[10$^{41}$ erg s$^{-1}$]\\
		\hline
		2023-10-15 06:52:00 & +11 & Swift-XRT &1.6&$<8 \times 10^{-3}$&$<$ 27.7&$<$ 173\\
		2023-10-20 16:45:00 & +16 & Swift-XRT &1.9&$<4 \times 10^{-3}$&$<$ 13.8&$<$ 86\\
		2023-10-25 09:14:00 & +21 & Swift-XRT &2.2&$<5 \times 10^{-3}$&$<$ 17.3&$<$ 108\\
		\hline
		2023-10-29 09:11:00 & +25 & Chandra &25 & $<5 \times 10^{-4}$ & $<$ 1.1 & $<$ 6.9\\
	\end{tabular}
	\caption{X-ray observations of SN 2023uqf, with limits in the 0.3--10\,keV band.  Phase is given relative to neutrino detection on 2023-10-04.}
	\label{tab:xray}
\end{table*}

\section{Radio Observations of SN 2023uqf}

Two epochs of X-band (8--12\,GHz) radio data were obtained on 2023 October 25 and 2023 December 19 under Program ID 23B-138 (PI: Ho) with the Karl G Janksy Very Large Array (VLA, \citep{vla}) while the array was in D configuration. 
3C286 was used as the flux density and bandpass calibrator and J0956+2515 as the complex gain calibrator.
Data were calibrated using the automated pipeline available in the Common Astronomy Software Applications (CASA; \citealt{McMullin2007}),
with additional flagging performed manually,
and imaged using the CLEAN algorithm \citep{Hogbom1974}.
No radio source was detected at the position of SN 2023uqf in either of our observations, which are summarised in Table \ref{tab:radio}.

\begin{table}[]
	\centering
	\begin{tabular}{c | c | c | c}
		Date & Phase & Central Frequency & Flux Limit \\
		\hline
		2023-10-25 & +21d & 10 GHz & $<$6\,$\mu$Jy \\
		2023-12-19 & +76d & 10 GHz & $<$6\,$\mu$Jy
	\end{tabular}
	\caption{Radio observations of SN 2023uqf at the VLA. Phase is given relative to neutrino detection.}
	\label{tab:radio}
\end{table}

\section{Neutrino Follow-up with ZTF}

In Table \ref{tab:neutrinos} we give details of all neutrino follow-up campaigns conducted by ZTF. This includes the 24 campaigns detailed in \citet{stein_23}, as well as 19 more recent campaigns conducted after December 2021. Coverage estimates from ZTF account for the small chip gaps between CCDs, with reduce our overall observed area. Our observations tiled a cumulative area of 257 sq. deg. across all campaigns.

\begin{longtable*}{c | c | c | c | c | c}%
	\hline%
	Neutrino&Full Area&ZTF Coverage&Signalness&Best{-}Fit Energy&Ref\\%
	Event&{[}sq. deg.{]}&{[}sq. deg.{]}&&{[}TeV{]}&\\%
	\hline%
	\endhead%
	\hline%
	IC190503A&1.95&1.69&0.36&/&\citep{IC190503A,IC190503A_ztf}\\%
	IC190619A&27.25&22.98&0.55&199&\citep{IC190619A,IC190619A_ztf}\\%
	IC190730A&5.44&4.87&0.67&299&\citep{IC190730A,IC190730A_ztf}\\%
	IC190922B&4.51&4.45&0.51&187&\citep{IC190922B,IC190922B_ztf}\\%
	IC191001A&25.64&20.73&0.59&217&\citep{IC191001A,IC191001A_ztf}\\%
	IC200107A&7.57&6.35&/&/&\citep{IC200107A,IC200107A_ztf}\\%
	IC200109A&22.45&20.11&0.77&375&\citep{IC200109A,IC200109A_ztf}\\%
	IC200117A&2.84&2.84&0.38&108&\citep{IC200117A,IC200117A_ztf}\\%
	IC200512A&9.72&9.57&0.32&109&\citep{IC200512A,IC200512A_ztf}\\%
	IC200530A&25.45&22.98&0.59&82&\citep{IC200530A,IC200530A_ztf}\\%
	IC200620A&1.74&1.36&0.32&114&\citep{IC200620A,IC200620A_ztf}\\%
	IC200916A&4.24&4.20&0.32&110&\citep{IC200916A,IC200916A_ztf}\\%
	IC200926A&1.77&1.74&0.44&670&\citep{IC200926A,IC200926A_ztf}\\%
	IC200929A&1.12&1.12&0.47&183&\citep{IC200929A,IC200929A_ztf}\\%
	IC201007A&0.56&0.56&0.88&683&\citep{IC201007A,IC201007A_ztf}\\%
	IC201021A&6.86&6.50&0.30&105&\citep{IC201021A,IC201021A_ztf}\\%
	IC201130A&5.46&5.37&0.15&203&\citep{IC201130A,IC201130A_ztf}\\%
	IC201209A&4.72&4.01&0.19&419&\citep{IC201209A,IC201209A_ztf}\\%
	IC201222A&1.53&1.12&0.53&186&\citep{IC201222A,IC201222A_ztf}\\%
	IC210210A&2.76&2.14&0.65&287&\citep{IC210210A,IC210210A_ztf}\\%
	IC210510A&4.04&3.55&0.28&113&\citep{IC210510A,IC210510A_ztf}\\%
	IC210629A&5.98&4.84&0.35&121&\citep{IC210629A,IC210629A_ztf}\\%
	IC210811A&3.17&3.14&0.66&218&\citep{IC210811A,IC210811A_ztf}\\%
	IC210922A&1.58&1.26&0.93&751&\citep{IC210922A,IC210922A_ztf}\\%
	IC220405A&8.44&8.34&0.32&122&\citep{IC220405A,IC220405A_ztf}\\%
	IC220405B&3.50&3.50&0.36&106&\citep{IC220405B,IC220405B_ztf}\\%
	IC220501A&3.56&2.93&0.40&127&\citep{IC220501A,IC220501A_ztf}\\%
	IC220513A&4.05&3.52&0.56&208&\citep{IC220513A,IC220513A_ztf}\\%
	IC220624A&9.78&8.14&0.61&193&\citep{IC220624A,IC220624A_ztf}\\%
	IC220822A&9.82&8.29&0.38&115&\citep{IC220822A,IC220822A_ztf}\\%
	IC220907A&5.74&5.29&0.46&128&\citep{IC220907A,IC220907A_ztf}\\%
	IC221216A&8.09&6.48&0.41&156&\citep{IC221216A,IC221216A_ztf}\\%
	IC221223A&1.48&1.10&0.79&353&\citep{IC221223A,IC221223A_ztf}\\%
	IC230112A&5.98&5.84&0.28&111&\citep{IC230112A,IC230112A_ztf}\\%
	IC230707A&1.73&1.52&0.66&280&\citep{IC230707A,IC230707A_ztf}\\%
	IC230724A&0.55&0.54&0.53&191&\citep{IC230724A,IC230724A_ztf}\\%
	IC230727A&2.60&2.40&0.29&113&\citep{IC230727A,IC230727A_ztf}\\%
	\textbf{IC231004A}&\textbf{4.29}&\textbf{3.59}&\textbf{0.84}&\textbf{442}&\citep{ic_gcn_1,ic_gcn_2,gcn_disc}\\%
	IC231103A&15.27&15.09&0.59&180&\citep{IC231103A,IC231103A_ztf}\\%
	IC231202A&8.80&8.66&0.27&108&\citep{IC231202A,IC231202A_ztf}\\%
	IC240105A&0.22&0.22&0.30&110&\citep{IC240105A,IC240105A_ztf}\\%
	IC240327B&7.79&7.54&0.37&153&\citep{IC240327B,IC240327B_ztf}\\%
	IC240721A&5.13&5.13&0.45&147&\citep{IC240721A,IC240721A_ztf}\\%
	\caption{Summary of all triggers by the ZTF neutrino follow-up program \citep{stein_23}, as of September 2024. IC190503A was released under the old `V1' IceCube alert schema, and did not have an energy estimate \citep{ic_realtime}. 
		IC200107A did not have a reported `signalness' or energy value \citep{IC200107A}, though for subsequent calculations we assume it has a `typical' signalness value of 0.5. IC231004A is highlighted in boldface.}
	\label{tab:neutrinos}
\end{longtable*}

\subsection{Probability of Chance Detection}
We estimate the probability of chance coincidence following the same procedure used in \citet{bran} and \citet{tywin}. As detailed in \citet{stein_23}, the vast majority ($>$98\%) of candidates identified by our ZTF program can ultimately be excluded after subsequent vetting and additional follow-up. Only a tiny fraction are consistent with the limited number of theoretically-motivated neutrino production scenarios which we search for (in particular supernova with CSM interaction, choked or successful relativistic jets, TDEs or flaring AGN) and the impact of the look-elsewhere effect is therefore relatively minimal.

As of September 2024, the ZTF Bright Transient Survey has identified 18 Type Ibn supernovae in a flux-limited sample down to a depth of m $=$ 18.5 since survey start in March 2018. Based on this rate, we estimate that 82 Type Ibn supernovae would have been discovered if the sample were complete down to m$_{g}$ $=$ 19.6 (the peak magnitude of SN 2023uqf) assuming a typical F$^{(-3/2)}$ scaling. 
However, the CCSN rate increases with redshift. We conservatively assume that all Ibn are as bright as SN 2023uqf, and calculate the relative CCSN rate increase between the BTS limit and our new 19.6 magnitude limit. We find that the number of expected supernovae would at most be 10\% higher. We correct for this factor, yielding a final expectation of 91 Type Ibn supernovae. 

These supernovae are distributed across the accessible extragalactic night sky for ZTF, which is approximately 28000 sq. deg., and a survey duration of 2389 days as of September 2024, giving a density of 1.37 $\times 10^{-6}$ new Type Ibn supernovae per sq. deg per day. Using the window of 8 days for which SN 2023uqf was detectable by ZTF, we thus estimate an overall density of SN 2023uqf-like sources at any given time to be 1.09 $\times 10^{-5}$ per sq. deg. of surveyed sky.

Accounting for all ZTF follow-up campaigns performed through to September 2024, which spanned a cumulative 257 sq. deg., the expected number of Ibn supernovae is 2.81$\times10^{-3}$. The probability that a Type Ibn supernova like SN 2023uqf would have been discovered by chance is thus 0.281\%, and is therefore disfavoured with a significance of 2.77$\sigma$. There is some uncertainty in this calculation, due to Poisson counting statistics. With 18 original supernovae, we expect a standard deviation of $\sqrt{N} = 4.24$, corresponding to $\pm 24\%$. Our chance coincidence probability is more precisely $0.28 \pm 0.06 \%$, and our statistical significance is thus $2.77 \pm 0.07 \sigma$.

We can also consider the broader population of interacting supernovae (including Types IIn and Icn as well as Type Ibn), for which an enlarged ZTF-detectable window of 70 days is more representative. The probability that our ZTF program detected any such supernova with evidence of CSM interaction is 12.6\%. The discovery of SN 2023uqf alone therefore does not provide strong evidence for neutrino production in the broader interacting supernova population. We would require additional neutrino-supernova associations before any conclusions about the entire interacting supernova population could be drawn, rather than just the rare subset of Type Ibn supernovae.

\subsection{Diffuse Neutrino Emission from Supernovae}
We use our sample of 43 neutrino follow-up campaigns to calculate the fraction of astrophysical neutrino alerts which arise from Type Ibn supernovae. 

We firstly consider sources which are bright enough to be detected by our program. From all Type Ibn supernovae which could potentially be detected by ZTF ToO observations, over the full course of our ZTF neutrino follow-up program, there has been exactly one observed Ibn-neutrino coincidence. This implies an implies an underlying expectation of 
$0.05 < N_{\nu, \textup{ZTF-Ibn}} < 4.74$ astrophysical neutrinos (at 90\% confidence) assuming Poisson counting statistics. 

We then divide this number by the total number of astrophysical neutrinos in the sample, to derive the fraction of neutrinos which arise from ZTF-detectable Type Ibn supernovae. Accounting for the astrophysical probability of each individual neutrino follow-up by ZTF, we expect a total of 20.61 astrophysical neutrinos from our 43 campaigns, and therefore expect that $0.24\% < f_{\nu, \textup{ZTF-Ibn}} < 14.6\%$ of astrophysical neutrinos arrive from ZTF-detectable Type Ibn supernovae. 

We emphasize that, while one might naively assume detected counterparts to neutrinos would be the closest examples of a population, this is very unlikely to be the case. Because neutrinos are detected individually, rather than from resolved sources, they are much less biased towards the local universe than traditional astronomy surveys. Though each individual source will be dimmer at higher redshift, this can be overcome by the greater volume contained within a redshift shell. For sources which traces the star formation rate, the median redshift for a neutrino source would be z$\approx$0.6, though this depends to some extent on the underlying neutrino spectrum shape (see e.g \citet{stein_23} for a more detailed discussion). Almost all ($\sim$95\%) of the neutrino flux from an SFR-like population would arise from sources beyond the `local' (z$<$0.1) universe, and our ZTF observations would only be sensitive to a fraction of all neutrino sources. It is only by conducting multiple campaigns that our follow-up program can be used to discover neutrino source populations. We expect that counterparts which are detected will likely be closer to the sensitivity limit of a given telescope.

We assume our ZTF program can recover sources with high efficiency out to a depth of 20.5 mag, which is realistic given both our historical performance \citep{stein_23} and the the typical depth of our ToO observations ($m > 21$). This depth would correspond to a detection threshold of z$=$0.22 for Type Ibn supernovae with an SN 2023uqf-like peak magnitude of $-19.7$ mag. By integrating the aggregate emission assuming that the supernova rate follows the underlying star formation rate, we estimate that these `detectable' supernovae out to z$=$0.22 should contribute 14.9\% of the total supernova neutrino flux.

We further assume that the intrinsic neutrino emission properties of the nearby Type Ibn supernovae are, on average, similar to those of higher-redshift supernovae. We divide our estimates by the completeness fraction, to measure the total flux from the Ibn population. We find that $1.61\% < f_{\nu, \textup{All Ibn}} < 97.98\%$ of the neutrino flux originates in the broader Type Ibn supernova population. Our follow-up program is thus sufficiently sensitive to constrain both the maximum and minimum contribution of this population to the diffuse neutrino flux, given our single observed Type Ibn-neutrino association. 

Previous IceCube analysis limited the aggregate emission of Type IIn supernovae to less than $<33.9\%$ neutrino flux \citep{ic_sn_23}. We note that the previous IceCube analysis also limited the contribution of stripped envelope supernovae to $<14.9\%$ of the total neutrino flux \citep{ic_sn_23}, but this limit is only valid under the assumption that neutrino emission follows an unbroken E$^{-2.5}$ power law, which is unlikely to be the case for e.g a choked jet scenario \citep[see e.g][]{murase_24}. Moreover, the IceCube limit is only valid under the assumption that all SESNe emit as neutrino `standard candles'. Given that only a small fraction ($\sim$6\%) of SESNe exhibit CSM interaction, this is unlikely to be the case. No dedicated neutrino source search has ever been conducted that targets those stripped-envelope supernovae which show CSM interaction. Instead, the most sensitive constraint on neutrino emission from these sources would ultimately be the ZTF neutrino follow-up program itself \citep{stein_23}. In any case, the previous constraints of neutrino emission from interacting supernovae subpopulations is fully consistent with the association presented here, which implies that at least 1.6\% of the astrophysical neutrino flux arises from interacting supernovae.   Complementary limits on neutrino emission from LFBOTs also exist, with no more than 16\% arising from this population \citep{stein_thesis}, based on the non-detection of neutrinos from AT 2018cow \cite{stein_icrc_19}.

\section{Photometry Table}

The full photometry for SN 2023uqf is given in Table \ref{tab:photometry}, including non-detections in the 30 days before neutrino detection. Beyond triggered follow-up, it also includes serendipitous coverage from the public ZTF \citep{bellm_19} and ATLAS \citep{atlas_18, atlas_transients_20} surveys. We note that ATLAS serendipitously observed the field of IC231004A in the minutes before and after neutrino detection, providing a useful constraint on bright contemporaneous optical emission to a depth of m=18.9.

\begin{longtable*}{c | c | c | c| c | c | c}
	\hline
	Time [UT]&Phase&Mag [AB]&$\Delta_{m}$& Limiting Mag & Filter&Instrument\\
	\hline
	\endhead
	\hline
	2023-09-15 12:19:37 & -19.1 d & / & / & 19.5 & r & ZTF\\
	2023-09-15 12:23:06 & -19.1 d & / & / & 19.4 & r & ZTF\\
	2023-09-15 12:26:34 & -19.1 d & / & / & 19.3 & r & ZTF\\
	2023-09-15 12:30:02 & -19.1 d & / & / & 19.1 & r & ZTF\\
	2023-09-15 12:33:30 & -19.1 d & / & / & 18.9 & r & ZTF\\
	2023-09-16 12:21:03 & -18.1 d & / & / & 19.6 & r & ZTF\\
	2023-09-16 12:24:35 & -18.1 d & / & / & 19.5 & r & ZTF\\
	2023-09-16 12:28:07 & -18.1 d & / & / & 19.4 & r & ZTF\\
	2023-09-16 12:31:36 & -18.1 d & / & / & 19.1 & r & ZTF\\
	2023-09-16 12:35:07 & -18.1 d & / & / & 18.7 & r & ZTF\\
	2023-09-17 12:24:12 & -17.1 d & / & / & 19.8 & r & ZTF\\
	2023-09-17 12:27:44 & -17.1 d & / & / & 19.5 & r & ZTF\\
	2023-09-17 12:31:15 & -17.1 d & / & / & 19.4 & r & ZTF\\
	2023-09-17 12:34:45 & -17.1 d & / & / & 19.2 & r & ZTF\\
	2023-09-17 12:38:16 & -17.1 d & / & / & 18.9 & r & ZTF\\
	2023-09-18 12:23:52 & -16.1 d & / & / & 19.7 & r & ZTF\\
	2023-09-18 12:27:24 & -16.1 d & / & / & 19.6 & r & ZTF\\
	2023-09-18 12:30:54 & -16.1 d & / & / & 19.4 & r & ZTF\\
	2023-09-18 12:34:25 & -16.1 d & / & / & 19.1 & r & ZTF\\
	2023-09-18 12:37:56 & -16.1 d & / & / & 18.8 & r & ZTF\\
	2023-09-19 12:26:04 & -15.1 d & / & / & 19.6 & r & ZTF\\
	2023-09-19 12:29:36 & -15.1 d & / & / & 19.5 & r & ZTF\\
	2023-09-19 12:33:08 & -15.1 d & / & / & 19.3 & r & ZTF\\
	2023-09-19 12:36:39 & -15.1 d & / & / & 19.0 & r & ZTF\\
	2023-09-20 12:23:10 & -14.1 d & / & / & 19.6 & r & ZTF\\
	2023-09-20 12:26:34 & -14.1 d & / & / & 19.6 & r & ZTF\\
	2023-09-20 12:29:59 & -14.1 d & / & / & 19.6 & r & ZTF\\
	2023-09-20 12:33:23 & -14.1 d & / & / & 19.6 & r & ZTF\\
	2023-09-20 12:36:48 & -14.1 d & / & / & 19.3 & r & ZTF\\
	2023-09-27 12:10:07 & -7.1 d & / & / & 20.0 & r & ZTF\\
	2023-09-27 12:13:39 & -7.1 d & / & / & 20.0 & r & ZTF\\
	2023-09-27 12:17:11 & -7.1 d & / & / & 19.9 & r & ZTF\\
	2023-09-27 12:20:43 & -7.1 d & / & / & 19.9 & r & ZTF\\
	2023-09-27 12:24:14 & -7.1 d & / & / & 20.0 & r & ZTF\\
	2023-09-28 12:12:06 & -6.1 d & / & / & 19.5 & r & ZTF\\
	2023-09-28 12:15:46 & -6.1 d & / & / & 19.6 & r & ZTF\\
	2023-09-28 12:19:26 & -6.1 d & / & / & 19.6 & r & ZTF\\
	2023-09-28 12:23:06 & -6.1 d & / & / & 19.6 & r & ZTF\\
	2023-09-28 12:26:45 & -6.1 d & / & / & 19.6 & r & ZTF\\
	2023-09-29 12:11:36 & -5.1 d & / & / & 19.3 & r & ZTF\\
	2023-09-29 12:15:16 & -5.1 d & / & / & 19.2 & r & ZTF\\
	2023-09-29 12:18:55 & -5.1 d & / & / & 19.2 & r & ZTF\\
	2023-09-29 12:22:35 & -5.1 d & / & / & 19.3 & r & ZTF\\
	2023-09-29 12:26:15 & -5.1 d & / & / & 19.3 & r & ZTF\\
	2023-09-30 14:41:49 & -4.0 d & / & / & 18.1 & o & ATLAS\\
	2023-09-30 14:49:16 & -4.0 d & / & / & 18.0 & o & ATLAS\\
	2023-09-30 14:55:09 & -4.0 d & / & / & 18.1 & o & ATLAS\\
	2023-09-30 15:20:47 & -4.0 d & / & / & 17.7 & o & ATLAS\\
	2023-10-02 14:20:07 & -2.0 d & / & / & 17.0 & o & ATLAS\\
	2023-10-02 14:24:45 & -2.0 d & / & / & 16.7 & o & ATLAS\\
	2023-10-02 14:33:40 & -2.0 d & / & / & 14.8 & o & ATLAS\\
	2023-10-02 14:57:23 & -2.0 d & / & / & 17.9 & o & ATLAS\\
	2023-10-04 14:25:36 & -0.0 d & / & / & 18.8 & o & ATLAS\\
	2023-10-04 14:31:04 & -0.0 d & / & / & 18.9 & o & ATLAS\\
	2023-10-04 15:04:11 & 0.0 d & / & / & 18.9 & o & ATLAS\\
	2023-10-04 15:13:11 & 0.0 d & / & / & 18.9 & o & ATLAS\\
	2023-10-05 11:54:57 & 0.9 d & 20.5 & 0.1 & 20.7 & g & ZTF\\
	2023-10-05 12:16:08 & 0.9 d & 20.5 & 0.1 & 21.0 & r & ZTF\\
	2023-10-06 11:50:22 & 1.9 d & / & / & 19.5 & g & ZTF\\
	2023-10-06 12:07:01 & 1.9 d & / & / & 19.7 & r & ZTF\\
	2023-10-07 11:46:23 & 2.9 d & 19.6 & 0.1 & 19.8 & g & ZTF\\
	2023-10-10 14:51:01 & 6.0 d & / & / & 15.4 & o & ATLAS\\
	2023-10-10 14:57:22 & 6.0 d & / & / & 16.7 & o & ATLAS\\
	2023-10-10 15:08:04 & 6.0 d & / & / & 17.4 & o & ATLAS\\
	2023-10-13 11:05:02 & 8.9 d & 20.0 & 0.2 & 20.0 & g & ZTF\\
	2023-10-13 11:21:39 & 8.9 d & 19.9 & 0.2 & 20.2 & g & ZTF\\
	2023-10-13 12:08:30 & 8.9 d & 20.0 & 0.1 & 20.4 & r & ZTF\\
	2023-10-13 12:16:48 & 8.9 d & 20.2 & 0.2 & 20.3 & r & ZTF\\
	2023-10-14 14:52:21 & 10.0 d & / & / & 19.3 & c & ATLAS\\
	2023-10-14 15:08:20 & 10.0 d & / & / & 19.3 & c & ATLAS\\
	2023-10-14 15:13:44 & 10.0 d & / & / & 19.3 & c & ATLAS\\
	2023-10-15 06:55:02 & 10.7 d & 19.5 & 0.1 & / & u & Swift\\
	2023-10-15 11:23:08 & 10.9 d & / & / & 19.9 & g & ZTF\\
	2023-10-15 11:42:01 & 10.9 d & 20.2 & 0.2 & 20.4 & r & ZTF\\
	2023-10-17 10:37:11 & 12.8 d & 20.6 & 0.1 & 22.0 & g & SEDM\\
	2023-10-17 10:43:10 & 12.8 d & 20.6 & 0.1 & 22.0 & r & SEDM\\
	2023-10-17 10:49:08 & 12.8 d & 20.7 & 0.1 & 21.5 & i & SEDM\\
	2023-10-18 10:43:33 & 13.8 d & / & / & 19.5 & u & SEDM\\
	2023-10-18 10:49:25 & 13.8 d & 20.6 & 0.1 & 21.7 & g & SEDM\\
	2023-10-18 10:55:17 & 13.8 d & 21.0 & 0.1 & 21.6 & r & SEDM\\
	2023-10-18 11:01:17 & 13.8 d & 20.7 & 0.1 & 21.4 & i & SEDM\\
	2023-10-19 10:40:24 & 14.8 d & 20.7 & 0.1 & 21.7 & g & SEDM\\
	2023-10-19 10:46:16 & 14.8 d & 21.1 & 0.1 & 21.6 & r & SEDM\\
	2023-10-19 10:52:09 & 14.8 d & 20.7 & 0.2 & 20.8 & i & SEDM\\
	2023-10-20 10:48:37 & 15.8 d & / & / & 21.3 & J & WIRC\\
	2023-10-21 10:46:07 & 16.8 d & 21.0 & 0.1 & 21.9 & g & SEDM\\
	2023-10-21 10:51:59 & 16.8 d & 21.5 & 0.2 & 21.5 & r & SEDM\\
	2023-10-21 10:57:52 & 16.8 d & 20.9 & 0.2 & 21.3 & i & SEDM\\
	2023-10-22 10:39:43 & 17.8 d & 21.1 & 0.1 & 22.0 & g & SEDM\\
	2023-10-22 10:45:38 & 17.8 d & 21.3 & 0.2 & 21.4 & r & SEDM\\
	2023-10-22 10:51:31 & 17.8 d & / & / & 16.5 & i & SEDM\\
	2023-10-25 10:07:11 & 20.8 d & / & / & 16.4 & g & SEDM\\
	2023-10-25 10:13:54 & 20.8 d & / & / & 17.7 & i & SEDM\\
	2023-11-01 04:35:02 & 27.6 d & 21.7 & 0.2 & / & r & ALFOSC\\
	2023-11-02 10:13:51 & 28.8 d & / & / & 20.9 & g & SEDM\\
	2023-11-02 10:20:07 & 28.8 d & / & / & 21.3 & r & SEDM\\
	2023-11-02 10:26:20 & 28.8 d & / & / & 20.7 & i & SEDM\\
	2023-11-02 12:08:33 & 28.9 d & / & / & 19.6 & i & ZTF\\
	2023-11-03 11:08:42 & 29.9 d & / & / & 19.8 & r & ZTF\\
	2023-11-03 12:25:17 & 29.9 d & / & / & 20.0 & g & ZTF\\
	2023-11-03 14:41:08 & 30.0 d & / & / & 22.5 & z & GMOS\\
	2023-11-07 11:45:35 & 33.9 d & 22.4 & 0.1 & 23.4 & g & LMI\\
	2023-11-07 12:00:21 & 33.9 d & 22.6 & 0.1 & 23.2 & r & LMI\\
	2023-11-07 12:03:16 & 33.9 d & / & / & 22.8 & i & LMI\\
	2023-11-08 11:27:56 & 34.9 d & / & / & 20.0 & g & ZTF\\
	2023-11-08 12:17:27 & 34.9 d & / & / & 19.4 & i & ZTF\\
	2023-11-10 11:40:01 & 36.9 d & 23.1 & 0.4 & 23.4 & g & WaSP\\
	2023-11-10 11:52:27 & 36.9 d & 22.8 & 0.4 & 23.3 & r & WaSP\\
	2023-11-10 12:03:24 & 36.9 d & / & / & 22.7 & i & WaSP\\
	2023-11-13 11:44:40 & 39.9 d & 22.4 & 0.1 & 24.8 & g & LMI\\
	2023-11-13 11:50:31 & 39.9 d & 22.8 & 0.1 & 25.0 & r & LMI\\
	2023-11-13 12:05:10 & 39.9 d & 23.4 & 0.1 & 24.3 & i & LMI\\
	2023-11-13 14:31:19 & 40.0 d & 22.6 & 0.3 & 22.9 & g & LCO-Sinistro\\
	2023-11-13 14:31:19 & 40.0 d & / & / & 22.4 & z & LCO-Sinistro\\
	2023-11-13 14:31:19 & 40.0 d & / & / & 22.5 & i & LCO-Sinistro\\
	2023-11-15 11:17:27 & 41.9 d & 23.4 & 0.4 & 23.5 & g & WaSP\\
	2023-11-15 11:56:27 & 41.9 d & / & / & 22.4 & i & WaSP\\
	2023-11-15 12:19:01 & 41.9 d & 23.1 & 0.4 & 23.3 & r & WaSP\\
	2023-12-08 12:37:56 & 64.9 d & / & / & 23.5 & g & WaSP\\
	2023-12-12 16:10:45 & 69.1 d & / & / & 23.8 & r & LCO-Sinistro\\
	2023-12-12 16:10:46 & 69.1 d & / & / & 23.3 & i & LCO-Sinistro\\
	2024-01-02 12:27:11 & 89.9 d & / & / & 21.3 & J & WIRC\\
	\caption{Photometric observations of SN 2023uqf. No magnitude is given in the case of non-detections or marginal detections below the nominal image depth, with the upper limit being given by the limiting magnitude. Phase is given relative to the time of neutrino detection.}
	\label{tab:photometry}
\end{longtable*}